\documentclass[utf8]{pepe} 

\setcitestyle{square} 
\usepackage{url,hyperref,lineno,microtype,subcaption}
\usepackage[onehalfspacing]{setspace}
\usepackage{epsfig}

\usepackage{hyperref}
\usepackage{float}
\usepackage{epsfig}
\usepackage{graphicx}
\usepackage{xcolor}
\usepackage{mathptmx}      
\usepackage{latexsym}
\usepackage{natbib}
\usepackage{longtable}
\usepackage{dcolumn}
\usepackage{comment}
\usepackage[utf8]{inputenc}

\DeclareMathVersion{nxbold}
\SetSymbolFont{operators}{nxbold}{OT1}{cmr} {b}{n}
\SetSymbolFont{letters}  {nxbold}{OML}{cmm} {b}{it}
\SetSymbolFont{symbols}  {nxbold}{OMS}{cmsy}{b}{n}

\DeclareMathAlphabet{\mathcal}{OMS}{cmsy}{m}{n}
\newcommand{\nk}{{\bf k}}

\newcommand{\nx}{{\bf x}}
\newcommand{\np}{{\bf p}}

\newcommand{\btau}{\vec \tau }
\newcommand{\vecq}{\mathbf{q}}
\newcommand{\vecP}{\mathbf{P}}
\newcommand{\vecp}{\mathbf{p}}
\newcommand{\vecr}{\mathbf{r}}

\def\keyFont{\fontsize{8}{11}\helveticabold}
\def\firstAuthorLast{Enrique Ruiz Arriola {et~al.}} 
\def\Authors{Enrique Ruiz Arriola\,$^{1,*}$
, Jose Enrique Amaro\,$^{1,*}$ and Rodrigo Navarro P\'erez\,$^{2}$}
\def\Address{$^{1}$Departamento de F\'{\i}sica At\'omica, Molecular y Nuclear and
  Instituto Carlos I de F{\'\i}sica Te\'orica y Computacional, \\
  Universidad de Granada E-18071 Granada, Spain.
\\$^{2}$ Department of Physics. San Diego State University. 5500 Campanile Drive, San Diego, California 02182-1233, USA}
\def\corrAuthor{Corresponding Author}

\def\corrEmail{earriola@ugr.es}

\begin{document}
\onecolumn
\firstpage{1}

\title[NN scattering and nuclear uncertainties]{NN scattering and
  nuclear uncertainties}

\author[\firstAuthorLast]{\Authors} 
\address{} 
\correspondance{} 

\extraAuth{}

\maketitle

\begin{abstract}


Ab initio calculations in Nuclear physics for atomic nuclei require a
specific knowledge of the interactions among their constituents,
protons and neutrons.  In particular, NN interactions can be
constrained down to scale resolutions of $\Delta r \sim 0.6 {\rm fm}$
from the study of phase shifts below the pion production
threshold. However, this allows for ambiguities and uncertainties
which have an impact on finite nuclei, nuclear- and neutron-matter
properties. On the other hand the nuclear many body problem is
intrinsically difficult and the computational cost increases with
numerical precision and number of nucleons.  However, it is unclear
what the physical precision should be for these calculations. In this
contribution we review much of the work done in Granada to encompass
both the uncertainties stemming from the NN scattering database in
light nuclei such as triton and alpha particle and the numerical
precision required by the solution method.

\tiny \keyFont{ \section{Keywords:} Nucleon-nucleon interaction,
  Scattering data, Uncertainty quantification, Nuclear Binding,
  Effective interactions, Statistical analysis}
\end{abstract}

\section{Introduction}

One of the main goals in Theoretical Nuclear Physics for many years has been
to achieve a sufficiently accurate {\it ab initio} solution of the Nuclear
Many Body Problem from a reductionist perspective. Within the present context
this means starting with the forces among the hadronic constituents, protons
and neutrons, and solving the corresponding quantum mechanical problem. While
this has been widely and openly recognized as an extremely difficult problem,
it already represents a simplification as compared to the fundamental problem
where the constituents are quarks and gluons building the nucleons and the
interactions are deduced from the gauge principle in QCD. The nuclear problem
schematically comprises two main steps i) the determination of the basic
interactions from spectroscopy and reactions at the few body level and ii) a
precise method of solution of the inferred interactions for the many body
problem. The predictive power of the theory corresponds therefore to the
relation between the input (nuclear two-, three-, four-body, and so on,
forces) and the output nuclear binding energies, form factors and nuclear
reactions, and the corresponding uncertainties.

The seminal paper of Yukawa~\cite{Yukawa:1935xg} established the first
theoretical evidence that the nuclear force has a finite range by the particle
exchange mechanism. The first determination of the tensor force and its
consequences for the deuteron were analyzed by Bethe~\cite{Bethe:1940iba,
Bethe:1940zz}.  The first $\chi^2$ statistical analyzes of NN scattering data
below pion production threshold started in the mid fifties~\cite{Stapp:1956mz}
(an account up to 1966 can be traced from Ref.~\cite{arndt1966chi}). A
modified $\chi^2$ method was introduced~\cite{perring1963nucleon} in order to
include data without absolute normalization.  The steady increase along the
years in the number of scattering data with better precision generated
incompatibilities and hence different criteria had to be
introduced~\cite{arndt1966determination, MacGregor:1968zzd, Arndt:1973ec} to
discard inconsistent data. For a  comprehensive review up to 1977 see
\cite{MacGregor:1960vw, Breit:1962zz, signell1969nuclear, deSwart:1977vpi}.
For a historical presentation before 1989 we recommend
Ref.~\cite{Machleidt:1989tm}.

Error analysis of NN phase-shifts for several partial waves became first
possible when the Nijmegen group~\cite{Stoks:1993tb} carried out a partial
wave analysis (PWA) fitting about 4000 experimental np and pp data, after
rejecting about 1000 inconsistent data with a $3\sigma$ criterion. The
analysis resulted in a value $\chi^2/{\rm dof} \sim 1$. In the fit the
potential was an energy dependent square well of radius $1.4 {\rm fm}$, plus 
one-pion-exchange (OPE) and charge-dependent (CD) contributions starting at
$1.4 {\rm fm}$, and a one-boson-exchange (OBE) piece operating below 2-2.5 fm.
Unfortunately, the required energy dependence becomes messy for nuclear
structure calculations.  In the next decade a variety of NN (energy
independent) potentials appeared in the literature fitting a large body of
scattering data with $\chi^2/{\rm dof} \sim 1$ \cite{Stoks:1993tb,
Stoks:1994wp, Wiringa:1994wb, Machleidt:2000ge, Gross:2008ps}, but
surprisingly error estimates on potential parameters were not made. While all
these modern potentials share the {\it local} OPE and CD tail and include
electromagnetic effects, the unknown short range components of these
potentials display a variety of forms and shapes: local
potentials~\cite{Stoks:1994wp}, nonlocal ones with angular momentum
dependence~\cite{Wiringa:1994wb}, energy dependence~\cite{Stoks:1993tb} or
momentum dependence~\cite{Stoks:1994wp, Machleidt:2000ge, Gross:2008ps}. While
in principle $p-$, $L-$ and $E-$non-localities are equivalent on-shell (see
e.g. Ref.~\cite{Amghar:1995av} for a proof in a $1/M_N$ expansion) they
reflect truly different physical effects and generally one should consider
them as independent quantities. Any specific choice results in a bias and
hence becomes a source of systematic errors.

Error propagation from nucleon-nucleon data to three- and four-nucleon
binding energies was pioneered in Ref.~\cite{Adam:1993pya}. A
rudimentary method based on coarse grained NN interactions was
proposed~\cite{NavarroPerez:2012vr,Perez:2012kt} providing a first
guess for error on bindings in nuclei and neutron and nuclear
matter. The Granada analysis of the triton using hyper-spherical
harmonics method was performed in \cite{Perez:2014laa}.  The triton
and the alpha particle were analyzed by solving the Faddeev equations
for $^3$H and the Yakubovsky equations for $^4$He in
\cite{Perez:2016oia}, and in {\em ab initio} no-core full
configuration calculations \cite{Perez:2015bqa}. Theoretical
uncertainties in the elastic nucleon-deuteron scattering observables
were calculated in\cite{Skibinski:2018dot},

While the history of the NN force and its applications to nuclear physics is
rather long, uncertainty quantification has not been addressed seriously until
recently (see e.g.~\cite{Dobaczewski:2014jga} for a review prefacing a full
volume of the ISNET community). There are several reasons why we think that
stressing this aspect of the theory may be particularly useful and fruitful.
One obvious one is to provide sensible error estimates in the theoretical
calculations. The traditional way was to try out several schemes and compare
the different results.  Another, less obvious reason, is to address the many
body nuclear problem within the realistic physical accuracy, rather than the
computational accuracy as it has been the customary approach up to now. This
applies in particular to the a priori accuracy of the solution of the nuclear
many body problem, which may eventually be relaxed as to facilitate
calculations not addressed before.  However, this may occur at a high price;
it is not unthinkable that any realistic attempt to quantify the theoretical
uncertainties may end up with a lack of predictive power on the side of the
theory.

We distinguish as usual in error analyses two sources of
uncertainties: statistical errors stemming from the data uncertainties
for a {\it fixed} form of the potential, and systematic errors arising
from the {\it different} most-likely forms of the potentials. Assuming
they are indenpendent, the total uncertainty corresponds to adding
both uncertainties in quadrature. In what follows it is advantageous
to take the viewpoint of considering any of the different potentials
as an independent but possibly {\it biased} way to determine the
scattering amplitudes and/or phase-shifts.  Because the biases
introduced in all single potential are independent on each other, a
randomization of systematic errors makes sense.

A prerequisite for such an analysis is to discern as much as possible
between statistical and systematic uncertainties. The former
correspond to the proper propagation of the experimental input while
the latter is concerned with the model or scheme dependence of the
calculation procedure. Systematic errors may include the genuine bias
to describe the physics and truncation errors which are related to the
approximate way the calculation is carried out. At the present stage,
the model bias is the largest source of uncertainty. 

After many years of tremendous efforts and steady progress, state of the art
calculations suggest that considerable success can be expected if one includes
the current knowledge of the two-, three-body forces and a variety of many
body techniques are applied. Going beyond four-body forces has never been
tried out, partly because of technical difficulties but also because of the
appearance of $\alpha-$clustering, based on the large stability and
compactness of the $^4$He nucleus, suggests that five body forces are
marginal~\footnote{Actually there are no purely contact interactions beyond
four body ones for fields with $(n,p,\uparrow, \downarrow)$ degrees of
freedom.}.

As already said, a credible quantification of the accuracy of the
theory requires a judicious determination of all sources of error in
the final results, including both the experimental information needed
to pin down the interactions as well as the convergence of the
numerical procedure used to solve the many body problem.  Given the
formidable computational effort needed to implement accurately many
body calculations ---even for light nuclei--- an {\it a priori}
determination of the errors induced from input data would very
helpful. This would set an useful accuracy goal and a limit beyond
which all refinements in the numerics would not improve the {\it
  theoretical} accuracy of the output. The purpose of the present work
is to review estimates on such limiting accuracy based on the
imperfect knowledge of the basic two body interactions.

Unfortunately, the situation we face in strong interactions in general
and in nuclear physics in particular is to compare and validate
inaccurate theories on the basis of accurate data. No theoretical
predictions outperforming experimental measurements in accuracy are
easily found. To make our point and concern more clear let us take for
instance the case of nuclear binding energies from a {\it
  semi-empirical} point of view, where a direct reference to nuclear
forces is mostly avoided. Bindings are experimentally known to high
accuracy, $\Delta B= 0.01-10$ KeV, whereas liquid-drop model inspired
mass fit formulas yield a lower theoretical accuracy $\Delta B= 0.6$
MeV (see e.g.  Refs.~\cite{Toivanen:2008im} and references
therein). This suggests that already within such a simple picture the
phenomenological theory is generally {\it not expected} to be more
accurate in its predictions than experiment.  Actually, according to
the standard $\chi^2/{\rm dof} \sim 1$ criterion the previous results
show that the theory is literally {\it incompatible} with data, and
thus not even an error analysis based on uncertainty propagation may
be undertaken. The situation is presumably less optimistic for the
{\it ab initio} approach based entirely on the knowledge of
(multiparticle) nuclear forces and a skillful solution of the nuclear
many body problem. This provides a motivation to quantify the accuracy
needed to solve the many body problem.

\section{Statement of the problem}

Let us be more specific on the meaning of uncertainty quantification in
nuclear physics. From a Hamiltonian describing A-nucleons, $H_A$, with kinetic
energy $T=\sum_{i=1}^A p_i^2/2M$ and multi-nucleon forces $V_{nN}$,
\begin{equation}
H_A = T + V_{2N} + V_{3N} + V_{4N}+ \dots \, ,   
\end{equation}
where
\begin{equation}
  V_{2N} = \sum_{i<j} V_{ij} \, , \quad V_{3N} = \sum_{i<j<k} V_{ijk} \, , \quad
  V_{4N} = \sum_{i<j<k<l} V_{ijkl} \, \quad \dots 
\end{equation}
one proceeds to solve the Schr\"odinger equation 
\begin{equation}
H_A\Psi_n = E_{n,A} \Psi_n \, .  
\end{equation}
In the absence of useful and accurate QCD-{\it ab initio} determinations,
phenomenological $V_{2N}$ interactions are {\it adjusted} to NN scattering
data and the deuteron, $^2$H ($A=2$), binding energy, while  $V_{3N}$ enter
into the $^3$H and $^3$He ($A=3$), bindings, $V_{4N}$ in $^4$He (A=4)  and so
on. Thus, the theoretical predictive power flow is expected to be from light
to heavy nuclei. For instance, in the case of the binding energy the problem
of error propagation based on NN force variations corresponds to
\begin{equation}
  V_{NN} = \bar V_{NN} \pm \Delta V_{NN} \to E_n(A)=\bar E_n(A) \pm
  \Delta E_n(A)
\end{equation}
The meaning of the variation $\Delta V_{NN}$ is a bit subtle, since there are
variations which are (scattering) equivalent and hence do not change the
scattering observables.

We are interested firstly in the NN scattering
problem~\cite{landau2008quantum}.  Quite generally we will consider
non-relativistic scattering of two particles with masses $m_1$ and $m_2$ where
$H=H_0+V$ and $H_0=p^2/2\mu$ is the kinetic energy and $\mu= m_1 m_2
/(m_1+m_2)$ the reduced mass (we drop ``$NN$'' for simplicity). The S-matrix
is defined as a boundary condition problem for $E \ge 0$
\begin{equation}
  S(E+i \epsilon)= 1 - 2\pi i \delta(E-H_0) T(E+ i \epsilon)
\end{equation}
where we have introduced the $T$-matrix which satisfies the scattering equation
in operator form,
\begin{equation}
  T(E)= V + V G_0(E) T(E) = V + VG_0(E)V + \dots = V ( 1- G_0 (E)
  V)^{-1}
  \label{Eq:LS}
\end{equation}
where in the second equality we write the exact summation of the
perturbative series. Other (complex) energy values are defined by
analytical continuation.  The T-matrix satisfies the reflection
property $T(E+i \epsilon)^\dagger = T(E-i \epsilon) $ if $V=V^\dagger$
in Eq.~(\ref{Eq:LS}) and hence the unitarity condition, $ S(E+ i
\epsilon) S(E+i \epsilon)^\dagger = 1 $, follows also from
$V=V^\dagger$ in Eq.~(\ref{Eq:LS}). The phase-shift is defined in
terms of the eigenvalues of the S-matrix, so that $ S \varphi_\alpha =
e^{2 i \delta_\alpha}\varphi_\alpha $ and for rotational invariant interactions
(we neglect spin to ease the notation) the scattering amplitude $M(\np', \np)$
is given by 
\begin{eqnarray}
M(\np', \np) = \sum_{lm} 4 \pi Y_{lm} (\np)
Y_{lm} (\np') \frac{e^{i \delta_l (p)} \sin \delta_l(p)}{p}= 
-\frac{2\mu}{4\pi} \langle \vec p'| T(E+i \epsilon) | \vec p\rangle
\Big|_{E_p=E_{p'}=E} 
\label{eq:pwe}
\end{eqnarray}
with $Y_{lm}(\np)$ the spherical harmonics and in our convention
$d\sigma/d\Omega =|M(\np', \np)|^2$ the differential cross
section. Any $NN$ unitary transformation, $U$, transforms the
Hamiltonian and hence the potential as $V \to \tilde V = U H U^\dagger
-H_0$. For an infinitesimal transformation $U= 1 + i \eta + \dots $,
where $\eta$ is a small self-adjoint two-body operator, the scattering
equivalent variation corresponds to the change $ \Delta V = i [ \eta,
  H] $. To see the effect on scattering, start with the LS equation in
the form $T^{-1}=V^{-1}-G_0$ which upon a variation of the potential
produces a variation of the T-matrix $\Delta T = T V^{-1} \Delta V
V^{-1} T$ and after some manipulation one gets
\begin{equation}
-i  \Delta T = ( 1 + T G_0)
\eta G_0^{-1}- G_0^{-1} \eta (1+G_0 T)
\end{equation}
so that sandwiching this expression between plane waves gives
\begin{equation}
  \Delta \langle \vec k' | T(E+i\epsilon) | \vec k\rangle =
  -i (E-E_{k'}+i\epsilon) \langle k' |\eta (1+ G_0 T) | k \rangle
  +i (E-E_{k}+i\epsilon) \langle k' |(1+T G_0) \eta  | k \rangle 
\end{equation}
which vanishes in the on-shell limit $E_k=E_{k'}=E$ and $\epsilon \to
0$.  Thus,
\begin{equation}
\Delta V = i [ \eta, H] \implies \Delta \langle \vec k' | T(E+i\epsilon) | \vec k\rangle \Big|_{E_k=E_{k'}=E} =0
\end{equation}
or equivalently for finite unitary transformations, using
Eq.~(\ref{eq:pwe}), $\delta_{l,H}(p) = \delta_{l,U H U^\dagger}(p) $.

Given this general ambiguity the long lasting problem has been to decide which
is the proper representation of the NN interaction based on NN scattering
data. This is in essence the so-called inverse scattering problem which has
been studied extensively in the past (see e.g. Refs.~\cite{chadan2012inverse,
newton2012inverse} for reviews) and requires additional strong assumptions to
fix the particular form of the potential. For instance, assuming a {\it local}
potential and {\it complete } knowledge of the phase-shifts in each partial
wave it is possible to determine the solution uniquely provided the binding
energies and long distance behavior of the corresponding bound states wave
functions allocated by the potential are known. Clearly, these inverse
scattering ambiguities have an impact on the solution of the many body
problem, as was documented long time ago in nuclear
matter~\cite{Coester:1970ai} and in the triton and alpha
particles~\cite{Tjon:1975sme}, just to mention two prominent examples (see 
Ref.~\cite{srivastava1975off} for a review).

Much of the arbitrariness is reduced by invoking an underlying theoretical
description in terms of hadronic degrees of freedom, which allows to compute
$V_{NN} (\vec x)$ in terms of one-,  two-,\ldots, pion exchanges.  which in
turn may be related to the $\pi N$ scattering process, involving coupling
constants for vertex interactions.  At present such a picture seems to hold
down to NN separations of about the elementary radius, $r_c=1.8$fm, below
which composite and finite size effects start playing a role That means that,
essentially, variations of the NN potential of are restricted  at least to $
\Delta V_{NN} (\vec x) = 0 $ for $ r \ge r_c \approx 1.8 {\rm fm} $.

\section{The NN potential}

\subsection{The concept of a potential}

In order to properly formulate the uncertainties of the potentials it would be
adequate to review first the meaning of a potential in nuclear physics. This
is of utmost importance but also intriguing. On the one hand the potential is
not an observable but on the other hand to our knowledge it is not practical
to carry out {\it ab initio} calculations in Nuclear Physics at the hadronic
level without potentials. Ultimately, one hopes to be able to provide a direct
link between the uncertainties in the input data and propagate them to the
output of the many body problem. As said, this is only possible by using
non-observable nuclear potentials as intermediate steps.

From a classical (and macroscopic) point of view, potential and force
can be measured directly by just determining the separation static
energy between two infinitely heavy sources. Such a definition admits
a direct extension to the quantum mechanical microscopic case and
specifically to the NN interaction assuming interpolating composite
local nucleon fields made out of three quarks. In essence, this is the
approach followed in recent years in lattice QCD where many of the
traditionally assumed features of the NN interaction seem to be
confirmed~\cite{Aoki:2009ji, Aoki:2011ep, Aoki:2013tba}. A major
drawback of this approach is that such a calculation determines the
static NN energy which would become a physical observable if nucleons
were infinitely heavy. The quantum mechanical problem needs adding
kinetic energy contributions. Moreover, the fact that low energy NN
scattering provides unnaturally large cross sections corresponds to an
extreme fine tuning which is beyond the present lattice capabilities.

\subsection{The tensorial structure}
\label{sec:tensNN}

Assuming isospin invariance for the moment, the most general form of the NN
interaction can be written as \cite{okubo1958velocity}
\begin{align} 
V({\vecp}~', \vecp) &  = 
 \:\, V_C \:\, + \btau_1 \cdot \btau_2 \, W_C +   
\left[ \, V_S \:\, + \btau_1 \cdot \btau_2 \, W_S 
\,\:\, \right] \,
\vec\sigma_1 \cdot \vec \sigma_2
-i \vec S \cdot (\vecq \times \vecP)  
\left[ \, V_{LS} + \btau_1 \cdot \btau_2 \, W_{LS}    
\right] \,
\nonumber \\ &+ 
\left[ \, V_T \:\,     + \btau_1 \cdot \btau_2 \, W_T 
\,\:\, \right] \,
\vec \sigma_1 \cdot \vecq \,\, \vec \sigma_2 \cdot \vecq  + 
\left[ \, V_{Q} + \btau_1 \cdot \btau_2 \, 
      W_{Q} \, \right] \,
\vec\sigma_1\cdot(\vecq\times \vecP\,) \,\,
\vec \sigma_2 \cdot(\vecq\times \vecP\,)
\nonumber \\ &+ 
\left[ \, V_{P} + \btau_1 \cdot \btau_2 \, 
      W_{P} \, \right] \,
\vec\sigma_1\cdot \vecP \,\,
\vec \sigma_2 \cdot \vecP\,
\, ,
\label{eq_nnamp}
\end{align}
where ${\vecp}\,'$ and $\vecp$ denote the final and initial nucleon momenta in
the CMS, respectively. Moreover, $\vecq = {\vecp}\,' - \vecp$ is the momentum
transfer, $\vecP =({\vecp}\,' + \vecp)/2$ the average momentum, and $\vec S
=(\vec\sigma_1+ \vec\sigma_2)/2 $ the total spin, with $\vec \sigma_{1,2}$ and
$\btau_{1,2}$ the spin and isospin operators, of nucleon 1 and 2,
respectively.

The scalar functions appearing in the potential, Eq.~(\ref{eq_nnamp}), depend
on \emph{both} initial and final momentum $\vecp$ and $\vecp'$ respectively.
Because of rotational invariance we may thus form three independent
invariants, such as $p,p'$ and also $\vecq \cdot \vecP$ (which vanishes
on-shell). Transforming  to coordinate space in the variable $\vecr$,
conjugate to $q$, we have
\begin{equation}
  V(\vecr, \vecP) = \int \frac{d^3 q}{(2\pi)^3}
  e^{i \vecq \cdot \vecr } 
\textstyle
\langle \vecP + \frac12 \vecq | V | \vecP - \frac12 \vecq \rangle \, , 
\end{equation}
where we take $\langle \vecP + \frac12 \vecq | V | \vecP - \frac12 \vecq
\rangle \equiv V(\vecp',\vecp)$.  The case where these functions depend {\it
only} on the momentum transfer $\vecq=\vecp'-\vecp$ corresponds in coordinate
space to a {\it local} potential, $V(\vecr,\vecP)= V(\vecr)$. Local potentials
are appealing because they provide physical insight besides being directly
manageable by means of a Schr\"odinger equation in configuration space.
Moreover, attaching a field theoretical interpretation to the interaction,
locality must be satisfied by heavy and point-like elementary nucleons which
act as static sources, so that in this case the potential becomes the static
energy between nucleons which is an unique observable defined by
\begin{equation}
E_{NN}(r) = V_{NN} (r) + 2 M_N + \mathcal{O} ( M_N^{-1})  ,
\end{equation}
where we assume $M_N \gg m_\pi, E $. Non-localities are expected to be weak 
because $P/M_N \ll 1$, and should have a larger influence at short distances
(see e.g. Ref.~\cite{Piarulli:2014bda} for an explicit implementation). The
finite mass effects generate some ambiguity in the definition of the potential
and, as we will see, are the largest source of uncertainties in nuclear
physics. In any case, there is some freedom that can be used advantageously to
{\it choose} ---by means of suitable unitary
transformations~\cite{Amghar:2002pf}--- a convenient form of the potential to
simplify the solution of the two-body problem, and to simplify a particular
scheme of the many body problem. We remind, however, that this choice may be a
source of bias and hence of systematic uncertainty.

\subsection{Operator basis}
\label{sec:operator}

In our analysis we will be using potentials which become local in the
partial wave basis. While the use of local potentials is very
appealing since the whole analysis simplifies tremendously, the truth
is that their use at all distances is questionable for extended
particles. However, the range of non-locality is determined by the
interaction and our analysis (see below) supports that on a scale
$\Delta r \sim 0.6 {\rm fm}$ non-locality is not essential.

The potential is written as a sum of functions multiplied by each operator
\begin{equation} 
  V(r) = \sum_{n=1,23} V_n(r) O^n
\end{equation}
The first 14 operators are charge independent and correspond to
the ones used in the Argonne $v_{14}$ potential
\begin{align}
  O^{n=1,14} =& 1, {\tau}_{1}\!\cdot\! {\tau}_{2},\,
  {\sigma}_{1}\!\cdot\! {\sigma}_{2}, ( {\sigma}_{1}\!\cdot\!
  {\sigma}_{2}) ( {\tau}_{1}\!\cdot\! {\tau}_{2}),\, S_{12}, S_{12}(
  {\tau}_{1}\!\cdot\! {\tau}_{2}),\, \nonumber \\ & {\bf
    L}\!\cdot\!{\bf S}, {\bf L}\!\cdot\!{\bf S} ( {\tau}_{1}\!\cdot\!
  {\tau}_{2}), L^{2}, L^{2}( {\tau}_{1}\!\cdot\! {\tau}_{2}),\, L^{2}(
  {\sigma}_{1}\!\cdot\! {\sigma}_{2}), \nonumber\\ &  L^{2}(
  {\sigma}_{1}\!\cdot\! {\sigma}_{2}) ( {\tau}_{1}\!\cdot\!
  {\tau}_{2}),\, ({\bf L}\!\cdot\!{\bf S})^{2}, ({\bf L}\!\cdot\!{\bf
    S})^{2} ( {\tau}_{1}\!\cdot\! {\tau}_{2})\ . 
\end{align}
These 14 components are denoted by $c$, $\tau$, $\sigma$, $\sigma\tau$,
$t$, $t\tau$, $ls$, $ls\tau$, $l2$, $l2\tau$, $l2\sigma$, $l2\sigma\tau$,
$ls2$, and $ls2\tau$. The remaining CD operators are
\begin{align}
  O^{n=15,21} =& T_{12}, \, ( {\sigma}_{1}\!\cdot\! {\sigma}_{2})
  T_{12}\, , S_{12}T_{12},\, (\tau_{z1}+\tau_{z2})\ , \nonumber \\ &(
  {\sigma}_{1}\!\cdot\! {\sigma}_{2}) (\tau_{z1}+\tau_{z2})\ , L^{2}
  T_{12} , L^{2} ( {\sigma}_{1}\!\cdot\! {\sigma}_{2}) T_{12} \, .
  \nonumber \\ & {\bf L}\!\cdot\!{\bf S}T_{12}, ({\bf L}\!\cdot\!{\bf
    S})^{2} T_{12}
\end{align}
and are labeled as $T$, $\sigma T$,$tT$, $\tau z $,$\sigma \tau z$, $l2T$,
$l2\sigma T$, $ls T$ and $ls2 T$. The first five were introduced by Wiringa,
Stoks and Schiavilla in~\cite{Wiringa:1994wb}; the following two were included
in~\cite{Perez:2013jpa} to restrict CD to the $^1S_0$
partial wave by following certain linear dependence relations between $V_{T}$,
$V_{\sigma T}$, $V_{l2T}$ and $V_{l2\sigma T}$.  The last two terms are
required for the CD on the $^3P_0$, $^3P_1$ and $^3P_2$ partial
waves.  To incorporate CD on $P$ waves two more operators need
to be added to the basis we used previously getting a total of $23$ operators
$O^n$.

As in our previous analysis we set $V_{tT} = V_{\tau z} = V_{\sigma \tau z} =
0$ to exclude CD on the tensor terms and charge asymmetries. 
To restrict CD to the $S$ and $P$ waves parameters the
remaining potential functions must follow
\begin{align}
  48 V_{l2 T} &= -5 V_{T} + 3 V_{\sigma T} + 12 V_{ls T} - 48 V_{ls2 T} \\
  48 V_{\sigma l2 T} &=  V_{T} - 7 V_{\sigma T} + 4 V_{ls T} - 16 V_{ls2 T} 
\end{align}
The algebraic relation between the operator basis in momentum space and in
configuration space is explicitly given in Ref.~\cite{NavarroPerez:2019sfj}
and several examples are displayed.

\subsection{The long range contributions}

As mentioned above, the potential becomes an observable within a QFT setup for
infinitely heavy hadronic sources. For the finite mass case one may use
instead a perturbative matching procedure between a QFT with hadronic (and
electro-magnetic fields) fields and the quantum mechanical problem, which
should work at sufficiently long distances. The hadronic QFT calculable
contribution is separated into two pieces, the strong (pion exchange) piece
and the purely EM piece,
\begin{equation}
  V_{\rm QFT} =  V_{\pi}(r) + V_{\rm EM}(r) \, . 
\end{equation}
The CD-OPE potential in the long range part of the interaction
is the same as the one used by the Nijmegen group on their 1993
PWA~\cite{Stoks:1993tb} and reads
\begin{equation}
 V_{m, \rm OPE}(r) = f^2\left(\frac{m}{m_{\pi^\pm}}\right)^2\frac{1}{3}m\left[Y_m(r){\mathbf \sigma}_1\cdot\mathbf{\sigma}_2 + T_m(r)S_{1,2} \right]
\end{equation}
being $f$ the pion coupling constant, ${\mathbf \sigma}_1$ and ${\mathbf
\sigma}_2$ the single nucleon Pauli matrices, $S_{1,2}$ the tensor operator,
$Y_m(r)$ and $T_m(r)$ the usual Yukawa and tensor functions,
\begin{align}
 Y_m(r) &= \frac{e^{-m r}}{m r}, \nonumber \\
 T_m(r) &= \left(1+ \frac{3}{mr} + \frac{3}{(mr)^2} \right)\frac{e^{-m r}}{m r}. 
 \label{eq:Yukawa}
\end{align}
CD is introduced by the difference between the charged
$m_{\pi^\pm}$ and neutral $m_{\pi^0}$ pion mass by setting
\begin{align}
 V_{{\rm OPE},pp}(r) &= V_{m_{\pi^0},\rm OPE}(r), \nonumber \\
 V_{{\rm OPE},np}(r) &= -V_{m_{\pi^0},\rm OPE}(r)+ (-)^{(T+1)}2V_{m_{\pi^\pm},\rm OPE}(r).
 \label{eq:BreakIsospinOPE}
\end{align}
The neutron-proton electromagnetic potential includes only a magnetic moment
interaction
\begin{equation}
V_{\rm EM, np}(r) = V_{\rm MM, np}(r) = -\frac{\alpha \mu_n}{2M_{n} r^3}  \left( \frac{\mu_{p}S_{1,2}}{2 M_p}  + \frac{{\bf L}\!\cdot\!{\bf S}}{\mu_{np}}  \right),
\end{equation}
where $\mu_n$ and $\mu_p$ are the neutron and proton magnetic moments, $M_n$
the neutron mass, $M_p$ the proton one and ${\bf L}\!\cdot\!{\bf S}$ is the
spin orbit operator. The EM terms in the proton-proton channel include one and
two photon exchange, vacuum polarization and magnetic moment,
\begin{equation}
 V_{\rm EM, pp}(r) = V_{\rm C1}(r) + V_{\rm C2}(r) + V_{\rm VP}(r) + V_{\rm MM, pp}(r)
\end{equation}
where
\begin{eqnarray}
  V_{\rm C1}(r) &=& \frac{\alpha'}{r} \ ,     \\
  V_{\rm C2}(r) &=& -\frac{\alpha\alpha'}{M_{p} r^2} \ ,  \\
  V_{\rm VP}(r) &=& \frac{2\alpha\alpha'}{3\pi r} \int^{\infty}_{1} e^{-2m_{e}rx}\left(1+\frac{1}{2x^{2}}\right)
            \frac{\sqrt{x^{2}-1}}{x^{2}}dx  \ ,                 \\
  V_{\rm MM, pp}(r) &=& -\frac{\alpha}{4M^{2}_{p} r^3} \left[
                 \mu^{2}_{p}S_{1,2}  + 2(4\mu_{p}-1){\bf L}\!\cdot\!{\bf S}  \right].
\end{eqnarray}  
Note that these potentials are {\it only} used above $r_c = 3 {\rm fm}$ and
thus form factors accounting for the finite size of the nucleon can be set to
one.  Energy dependence is present through the parameter
\begin{equation}
  \alpha'= \alpha\frac{1+2k^2/M_p^2}{\sqrt{1+k^2/M_p^2}} \, , 
\end{equation}
where $k$ is the center of mass momentum and $\alpha$ the fine structure
constant. Table \ref{tab:Constants} lists the values used for the fundamental
constants in our calculations and typically used since the
benchmarking Nijmegen analysis.

\begin{table}[ht]
  \caption{\label{tab:Constants} Values of fundamental constants used.}
  \vskip.3cm
 \begin{tabular*}{\columnwidth}{@{\extracolsep{\fill}} c D{.}{.}{3.7} l}
   Constant     & \multicolumn{1}{c}{Value}      & Units      \\
  \hline  
   $\hbar c$     & 197.327053 & MeV fm    \\
   $m_{\pi^0}$   & 134.9739   & MeV$/c^2$   \\
   $m_{\pi^\pm}$ & 139.5675   & MeV$/c^2$   \\
   $M_{p}$       & 938.27231  & MeV$/c^2$   \\
   $M_{n}$       & 939.56563  & MeV$/c^2$   \\
   $m_e  $       & 0.510999   & MeV$/c^2$   \\
   $\alpha^{-1}$ & 137.035989 &           \\
   $f^2$         & 0.075      &           \\
   $\mu_p$       & 2.7928474  & $\mu_0$     \\
   $\mu_n$       &-1.9130427  & $\mu_0$     \\
 \end{tabular*}
\end{table}

\subsection{Short range contributions}

The short range contributions are fundamentally unknown and, despite
some lattice QCD efforts~\cite{Aoki:2009ji, Aoki:2011ep, Aoki:2013tba,
  Walker-Loud:2014iea}, can only be determined indirectly and
phenomenologically, mostly from NN scattering. Along the years some
experience has been gathered about the size, shape and range of the
potentials in the bulk, at least in configuration space, so that
refinements are made by a $\chi^2$ minimization to pp and np
scattering data (see below).  Besides, the analysis of scattering data
allows to obtain information on the lowest distance where the long
range contributions can be trusted. We anticipate that they may be
assumed to be valid for $r_c \ge 1.8$ fm when OPE and TPE
contributions are included. This coincides {\it a fortiori} with the
distance above which protons interact by Coulomb force as point-like
particles, and also with the typical distance between nucleons in
nuclear matter, $d= \rho^{-1/3}= 1.8 {\rm fm}$ for $\rho=0.17 \,{\rm
  fm}^{-3}$.

Finally, there is the issue on {\it which} and {\it how many} parameters are
needed to describe the short range force in a satisfactory manner. The primary
2013 Granada analysis has been carried out in terms of the so-called coarse
grained potentials~\cite{Perez:2013mwa}. The coarse grain procedure {\it
samples} the interaction with an optimal grain size, corresponding roughly to
the reduced de Broglie wavelength $\Delta r = \hbar /p$. For the maximum LAB
energy, 350 MeV, this corresponds to $\Delta r=0.6$ fm. Thus, we do not need
to sample the potential functions $V_i (r)$ at {\it all} points, but rather in
a grid of points, $V_i (r_n)$ given by $r_n = n \Delta r$.  We consider the
$V_i (r_n)$ values as fitting parameters. The particular interpolations
between these points are not physically relevant,  because shorter scales than
$\Delta r$ cannot be probed by the scattering process below a maximal $p = \sqrt{T_{\rm LAB} M_N /2} \sim 2{\rm fm}^{-1}$.

The number of grid points depends on the cut distance, $r_c$, above which the
functional form of the potential is known and corresponds to $N= r_c /\Delta
r$. Thus, the simplest case corresponds to $r_c=1.8$ fm and $N=3$ grid points
for any  radial component, $V_i(r_n)$, in the operator basis. In the partial
wave basis some refinements can be incorporated since the centrifugal barrier
limits the sampling points below the barrier in the classically forbidden
region, so that the estimate is~\cite{Fernandez-Soler:2017kfu,
RuizArriola:2019pwt},
\begin{equation}
  N_{\rm Par} \sim \frac12 (p_{\rm CM}^{\rm max} r_c)^2 \, g_S \, g_T \, , 
  \label{eq:npar}
\end{equation}
where $g_S$ and $g_T$ are spin and isospin degeneracy factors. The counting of
parameters for pp and np~\cite{Perez:2013cza} yields about 40 ``grained''
points $r_n$ in the fit carried up to a maximum energy $T_{\rm LAB} \le 350$
MeV. This {\it a priori} estimate coincides in the bulk with the number of parameters which
have traditionally been needed to fit data satisfactorily in the past. The
previous argument suggests that including more parameters is not expected to
improve significantly the fits to scattering data, but rather increase the
correlations among the $V_i(r_n)$ parameters.

There are many possible ways to describe the interaction at the ``grained''
points. The simplest is to consider Dirac delta-shells located at the sampled
points~\cite{Aviles:1973ee, Entem:2007jg}
\begin{equation}
V(r)|_{\rm Short} = \Delta r  \sum_{i,n} O_i V_i (r_n) \delta (r-r_n) \qquad r \le
r_c 
\end{equation}
We refer to Ref.~\cite{NavarroPerez:2011fm} for a pedagogical presentation of
coarse grained interactions which solve the Schr\"odinger equation by a 
discretized form~\cite{Aviles:1973ee, Entem:2007jg} of the variable phase
approach of Calogero~\cite{calogero1967variable}. This delta-shells
decomposition implies a similar one at the partial waves level, so that one
may use the partial wave strengths $V_{LL'}^{JS} (r_n)$ as fitting parameters.
This choice  is rather convenient for least squares minimization as the low
angular momentum partial wave components of the potential  are largely
uncorrelated, substantially speeding up the minimum
search~\cite{Perez:2014yla, Perez:2014kpa}. The transformation matrix from the
$V_i(r_n)$ to the $V_{LL'}^{JS} (r_n)$ basis can be found in
Ref.~\cite{Perez:2013jpa}.

\section{Partial Wave Analysis}

The NN scattering amplitude has five independent complex components which are
a function of energy and scattering angle~\cite{puzikov1957construction},
\begin{align}
M 
=& a
+ m (\mathbf{\sigma}_1\cdot\mathbf{n})(\mathbf{\sigma}_2\cdot\mathbf{n}) 
+ (g-h)(\mathbf{\sigma}_1\cdot\mathbf{m})(\mathbf{\sigma}_2\cdot\mathbf{m}) 
\nonumber \\ 
+& 
(g+h)(\mathbf{\sigma}_1\cdot\mathbf{l})(\mathbf{\sigma}_2\cdot\mathbf{l})  
+ c (\mathbf{\sigma}_1+\mathbf{\sigma}_2)\cdot\mathbf{n} \, . 
\end{align}
We use the three unit vectors ($\mathbf{k}_f$ and $\mathbf{k}_i$ are relative
final and initial momenta), 
\begin{equation}
\mathbf{l} =   \frac{\mathbf{k}_f+\mathbf{k}_i}{|\mathbf{k}_f+\mathbf{k}_i|} \, , 
 \qquad 
 \mathbf{m}=  \frac{\mathbf{k}_f-\mathbf{k}_i}{|\mathbf{k}_f-\mathbf{k}_i|} \, , \qquad 
 \mathbf{n}=  \frac{\mathbf{k}_f \wedge \mathbf{k}_i}{|\mathbf{k}_f \wedge\mathbf{k}_i|} \, . 
\end{equation}
For this amplitude the total spin $S$ is conserved and in this case the
partial wave expansion reads,
\begin{align}
  M^s_{m_s',m_s}(\theta) =& \frac{1}{2ik}
  \sum_{J,l',l}\sqrt{4\pi(2l+1)}Y^{l'}_{m_s'-m_s}(\theta,0) \nonumber
  \\ \times&
  C^{l',S,J}_{m_s-m_s',m_s',m_s}i^{l-l'}({\bf S}^{J,S}_{l,l'}-\delta_{l',l})
  C^{l,S,J}_{0,m_s,m_s},
 \label{eq:MmatrixPartialWaves}
\end{align}
where ${\bf S}$ is the unitary coupled channel S-matrix, and the $C's$ are
Clebsch-Gordan coefficients, $  C^{l,S,J}_{m,m_s,M}= \langle lmSM_s | JM \rangle$. The spins of the nucleon pair can be coupled to
total spin $S=0,1$ and hence $J=L \pm 1$ for unnatural parity, $(-1)^{L+1}$
states and $J=L$ for natural parity states. This amplitudes contains all
measurable physical information and the relation to observable quantities such
as differential cross sections and polarization asymmetries can be found in
Refs.~~\cite{Hoshizaki:1969qt, Bystricky:1976jr}. 

In the Stapp-Ypsilantis-Metropolis (SYM) representation \cite{Stapp:1956mz} 
the  $S$-matrix is written in terms of the \textit{nuclear-bar phase shifts}
$\bar \delta_{j\pm 1}$ and $\bar \epsilon_j$ as Denoting the phase shifts as
$\delta^{J,s}_{l,l'}$, for the singlet ($s=0$, $l = l'= J$) and triplet
uncoupled ($s=1$, $l=l'=J$) channels the $S$ matrix is simply
$e^{2i\delta^{J,s}_{l,l}}$, in the triplet coupled channel ($s=1$, $l=J\pm1$,
$l'=J\pm1$) it reads
\begin{equation}
S^J = \left( 
\begin{array}{c c}
 e^{2i\delta^{J,1}_{J-1}} \cos{2 \epsilon_J} & ie^{i(\delta^{J,1}_{J-1}+\delta^{J,1}_{J+1})} \sin{2 \epsilon_J} \\
 ie^{i(\delta^{J,1}_{J-1}+\delta^{J,1}_{J+1})} \sin{2 \epsilon_J} & e^{2i\delta^{J,1}_{J+1}} \cos{2 \epsilon_J}
\end{array}
\right),
\end{equation}
with $\epsilon_J$ the mixing angle.

The partial wave expansion provides an indirect way to find out the range of
nuclear forces by truncating the expansion. According to the standard
semi-classical argument (see e.g. \cite{Binstock:1972gx}), for an impact
parameter $b=(J+1/2)/p$ ($p$ is the CM momentum) the no-scattering condition
corresponds to $b \ge a$, so that $|\delta_{J_{\rm max}}| \le \Delta
\delta_{J_{\rm max}} $ where maximal angular momentum is provided by $J_{\rm
max} \approx p a$ with $a$ the range of the force. For the Yukawa OPE
interaction the exponential fall-off of the potential also means a similar
behavior for the phase-shifts, so typically one takes $S,P,D$ and $F$ waves
as {\it active} if the condition is $J+1/2 \approx p r_c $ with $r_c$ the
separation distance. 


We will review briefly the basics of scattering from a NN potential for
completeness and to provide our notation. Details may be found in standard
textbooks on scattering theory (see e.g. ~\cite{goldberger2004collision}). The
generalization of the well-known Rayleigh expansion for spin $S$ is
\begin{equation}
e^{i\nk\cdot\nx}\chi_{{}_{SM_s}}=
4\pi\sum_{l,m} i^l j_l(kr) Y^*_{l,m}(\hat{\nk}) \sum_{J,M}
\langle lmSM_s | JM \rangle \mathcal{Y}_{l S J M}(\hat{\nx}) \, ,
\label{Rayleigh-formula-spin}
\end{equation}
where $\chi_{{}_{SM_s}}$ is an eigenspinor with spin quantum numbers
$(S,M_s)$, and the functions $\mathcal{Y}_{l S J M}(\hat{\nx})$ are the
couplings of the spherical harmonics with the spinors $\chi_{{}_{SM_s}}$
to total angular momentum $J$,
\begin{equation}
\mathcal{Y}_{l S J M}(\hat{\nx}) =
\sum_{m^\prime,M^\prime_s}
\langle lm^\prime  SM^\prime_s | JM \rangle
Y_{l,m^\prime}(\hat{\nx})\, 
\chi_{{}_{SM^\prime_s}} \, .
\end{equation}
The local (but angular momentum dependent) NN potential described in the
previous section conserves spin $S$ and total angular momentum $J$, but not
the orbital angular momentum $L$. Therefore the scattering wave function for
spin $S$ is expanded as
\begin{equation}
  \Psi_{\nk, SM_s}(\nx)=
  4\pi\sum_{lmJM} i^l
   Y^*_{l,m}(\hat{\nk}) \langle lmSM_s | JM\rangle
\sum_{l'}
   \frac{u^{\,SJ}_{l'l}(r)}{kr}   
   \mathcal{Y}_{l' S J M}(\hat{\nx}) \, .
\label{partial-wave-bg}
\end{equation}
where the reduced radial wave functions $u^{\,SJ}_{l'l}(r)$ satisfy the
coupled channel differential equations
\begin{equation}
  \left[ -\frac{d^2}{dr^2}+ \frac{l'(l'+1)}{r^2} - k^2
    \right]
    u^{SJ}_{l'l}
    + \sum_{l''} U_{l',l''}^{SJ} (r) u^{SJ}_{l''l}=0    
\end{equation}
and the reduced potential is defined as $U(r) = 2\mu V(r)$. For regular
potentials the boundary condition at the origin reads
\begin{equation}
  u^{SJ}_{l'l} (r) \sim r^{l'+1} \qquad ( r \to 0 ) 
\end{equation}
The integration of the equations can advantageously be done using the delta
shell representation of the NN potential taking $\Delta r=0.6$ fm for $r \le
r_c$ (the coarse-grained and unknown part) and $\Delta r=0.1$ fm for
$r\ge r_c$ (the known field theoretical part). The complete set of
equations including Coulomb forces is provided in Ref.~\cite{Perez:2013jpa}.
The scattering boundary condition
\begin{equation}
\Psi_{S,m_s} (\vec x) \to e^{i \vec k \cdot \vec x} \chi_{S,m_s} +
\frac{e^{i k r}}{r}\sum_{m_{s'}=-S}^S M_{m_S,m_{S'}} \chi_{S,m_s'}
\end{equation}
implies a similar asymptotic condition for the reduced radial wave functions.
For the uncoupled case, $l=J$, one has for $r \sim R \gg 1/m_\pi$
\begin{equation}
u_J(r) \equiv u_{JJ}(r) \to \hat j_J(kr) - \cot{\delta_J(k)}\hat y_J(kr)
\end{equation}
where $\hat j_J(x) = x j_J(x)$ and $\hat y_J(x) = x y_J(x)$ are the
reduced spherical Bessel functions of order $J$ and $\delta_J =
\delta^{1 J}_J,\delta^{0 J}_J$. In the coupled triplet case, $S=1$,
the four wave functions $u_{l'l}(r)$, with $l',l=J-1,J+1$, are coupled
in pairs. The pair
\begin{equation}
  v_{\alpha J }
  = u_{J-1,J-1}  \kern 1cm  
  w_{\alpha J }
  = u_{J+1,J-1}
\end{equation}
verifies the coupled  equations
\begin{align}
  \left[ -\frac{d^2}{dr^2}+ \frac{J(J-1)}{r^2} - k^2
    \right]
    v_{\alpha J }
    + U_{J-1,J-1}^{SJ} (r)     v_{\alpha J }
    + U_{J-1,J+1}^{SJ} (r)     w_{\alpha J } &= 0
\label{coupled1} \\    
  \left[ -\frac{d^2}{dr^2}+ \frac{(J+1)(J+2)}{r^2} - k^2
    \right]
    w_{\alpha J }
    + U_{J+1,J+1}^{SJ} (r)     w_{\alpha J }
    + U_{J+1,J-1}^{SJ} (r)     v_{\alpha J } &= 0
\label{coupled2}    
\end{align}
On the other hand the pair
\begin{equation}
  w_{\beta J }
  = u_{J+1,J+1}  \kern 1cm  
  v_{\beta J }
  = u_{J-1,J+1}
\end{equation}
verifies the same coupled equations by changing $\alpha\rightarrow\beta$. This
is equivalent to say that the system (\ref{coupled1},\ref{coupled2}) has two
linearly independent solutions that we label as $\alpha$ and $\beta$
solutions. Their asymptotic behavior can be expressed in terms of the eigen
phase shifts as,
\begin{align}
v_{\alpha J}(r) \to & \hat j_{J-1} (kr) \cot \delta^{1J}_{J-1} - \hat y_{j-1} (kr)\\
w_{\alpha J}(r) \to &
\tan \epsilon_J
\Big[ \hat j_{j+1} (kr)\cot \delta^{1J}_{J-1} - \hat y_{j+1}(kr)\Big]
\\ 
v_{\beta J}(r) \to &
-\tan \epsilon
\Big[ \hat j_{j-1} (kr) \cot \delta^{1J}_{J+1} - \hat y_{j-1} (kr)\Big] \\
w_{\beta J}(r) \to & \hat j_{j+1} (kr)\cot \delta^{1J}_{J+1} - \hat y_{j+1}(kr)
\end{align}
This is known as the Blatt-Biedenharn (BB) parameterization in terms of the
{\em eigen phase shifts} $\delta^{1j}_{j\pm 1}$ and $ \epsilon_j$. These are
related to the nuclear-bar phase shifts  by the following equations
\begin{align}
  \delta^{1J}_{J-1} + \delta^{1J}_{J+1} =& 
  \bar \delta^{1J}_{J-1} + \bar \delta^{1J}_{J+1}\\
  \sin(\bar \delta^{1J}_{J-1} - \bar \delta^{1J}_{J+1}) =&
  \frac{\tan{2\bar \epsilon_J}}{\tan{2\epsilon_J}}\\
\sin(\delta^{1J}_{J-1} - \delta^{1J}_{J+1}) =& \frac{\sin{2\bar \epsilon_J}}{\sin{2\epsilon_J}}
\end{align}
Unless otherwise stated, in this work the phase shifts will always be
assumed to be the nuclear-bar ones. The Coulomb force is included
exactly by replacing in the previous formulas the Bessel functions
$j_l$ and $y_l$ by Coulomb functions $F_l$ and
$G_l$~\cite{goldberger2004collision}. The inclusion of magnetic
moments effect is complicated by their $1/r^3$ behaviour requiring
about 1000 partial waves~\cite{Perez:2013jpa}.

\section{Statistics}

The statistical treatment we follow here is quite standard, and we
list for the benefit the newcomer to the field the main steps to be
discussed in the following subsections. We first address the existing
scattering data and then we formulate the nature of the problem and
the standard $\chi^2$ approach searching for the most likely
potential. This requires discriminating between consistent and
inconsistent data, something which can be formulated in terms of a
self-consistent selection problem. After this, a direct statistically
satisfactory result can be deduced and, more importantly, error
propagation may legitimately be carried out in terms of the
corresponding covariance matrix implementing statistical
correlations. This allows in particular to determine the scattering
phase-shifts with error bars reflecting directly the experimental
uncertainties. More generally, it allows to transport these
experimental errors to any observable based on the nucleon-nucleon
potential. We will call these the statistical errors.

\subsection{Scattering data}

Once we have defined the potential model and the scattering formalism we may
proceed to determine the potential parameters $V_i (r_n)$ from the available
np and  pp scattering data and from the corresponding scattering observables
which are obtained from the scattering amplitude~\cite{Hoshizaki:1969qt,
Bystricky:1976jr} (see also tables~\ref{tab:chi2-obs-pp} and
~\ref{tab:chi2-obs-np} below for the notation). The compilation of the
existing published data since 1950 till 2013 is described in detail in
Ref.~\cite{Perez:2013jpa} and comprises 8124 fitting data including 7709
experimental measurements and 415 normalizations provided by the
experimentalists.

\subsection{Statement of the problem}

The finite amount, precision and limited energy range of the data as well as
the many different observables calls for a standard statistical $\chi^2$-fit
analysis~\cite{evans2004probability, eadie2006statistical}. This approach is
subjected to assumptions and applicability conditions that can only be checked
{\it a posteriori} in order to guarantee the self-consistency of the analysis.
Indeed, scattering experiments deal with counting Poisson statistics and
for moderately large number of counts a normal distribution is expected. 
Thus, one hopes that a satisfactory theoretical description $O_i^{\rm th}$ can
predict a set of N independent observed  data $O_i$ given an experimental
uncertainty $\Delta O_i$ as
\begin{equation}
O_i=O_i^{\rm th} + \xi_i \Delta O_i 
\label{eq:generator}
\end{equation}
with $i=1, \dots, N$ and $\xi_i$ are independent random {\it normal} variables
with vanishing mean value $\langle \xi_i \rangle =0$ and unit variance
$\langle \xi_i \xi_j \rangle = \delta_{ij}$, implying that $\langle O_i
\rangle =O_i^{\rm th} $. Establishing the validity of Eq.~(\ref{eq:generator})
is of utmost importance since it provides a basis for the statistical
interpretation of the error analysis.

\subsection{The least squares minimization}

If the $\xi_i$ are independent normal variables,then $\sum_{i=1}^\nu \xi_i^2$
represents a $\chi^2$ distribution with $\nu$ degrees of freedom. Thus, under
this {\it hypothesis} we may consider the standard $\chi^2$ method, which in
our case is defined as
\begin{equation}
  \chi^2 [ V_k(r_n) ]=
  \sum_{i=1}^{N_{\rm Dat}}
  \left[ \frac{O_i^{\rm exp} -O_i^{\rm th} ( V_k (r_n)) }{\Delta O_i^{\rm exp}}
    \right]^2   
\end{equation}
where $O_i^{\rm exp}$ is the experimental observable, $\Delta O_i^{\rm exp}$
its estimated uncertainty and $O_i^{\rm th} ( V_k (r_n))$ are the theoretical
results which depend on the fitting parameters $V_k (r_n)$, the values of the
potentials at the sampled points $r_n$. The least squares minimization has
always a solution which may be a global or a local minimum, namely 
\begin{equation}
  \chi^2_{\rm min} = \min_{V_k(r_n)} \chi^2 [ V_k(r_n) ] \equiv \chi^2
      [ \bar V_k(r_n) ]
\end{equation}
where $\bar V_k(r_n) $ the minimizing parameters. Basically, this minimization
eliminates $N_{\rm Par}$ parameters from the $N_{\rm Dat}$ data and we are
left with $\nu = N_{\rm Dat}-N_{\rm Par}$ degrees of freedom. The important
aspect here is the {\it statistical significance} of the procedure. This can
be checked {\it a posteriori} by looking at the residuals
\begin{equation}
R_i =  \frac{O_i^{\rm exp} -O_i^{\rm th}|_{\rm min}}{\Delta O_i^{\rm exp}} 
\end{equation}
where $ O_i^{\rm th}|_{\rm min} = O_i^{\rm th} ( \bar V_k
(r_n))$. According to the assumption underlying the $\chi^2$-method,
the set of variables $R_1 , \dots, R_{N_{\rm par}}$ should be
distributed as normal variables, i.e. they should look as $N_{\rm
  Par}$ variables extracted from a normal distribution $N(0,1)$.  For
a finite sample the veracity of this hypothesis can only be
established in probabilistic terms, so that we may estimate how likely
or unlikely would it be to accept of reject the starting normality
assumption.  Technically, this can be done in a variety of ways (see
e.g.  \cite{Perez:2014yla, Perez:2014kpa, Perez:2015pea}), but the
most popular measure of goodness of a fit is the $\chi^2$-test which
requires that the fit is accepted if
\begin{equation}
\frac{\chi^2_{\rm min}}{\nu} = 1 \pm \sqrt{\frac2{\nu}}
\end{equation}
with $\nu=N_{\rm Dat}-N_{\rm Par}$.  More elaborate tests may be
applied and we refer to \cite{Perez:2014yla, Perez:2014kpa,
  Perez:2015pea} for further details. In practice this means that for
$N_{\rm Dat}=8000$ and $N_{\rm Par}=50$ we should get $\chi^2_{\rm
  min}/\nu= 1 \pm 0.016 $ in order to validate
Eq.~(\ref{eq:generator}). Note that this is {\it very different} than
the loose claims in the literature where $\chi^2/\nu \approx 1$
qualifies for a good fit, complemented with a visual inspection of the
phase shifts. We emphasize that {\it looking similar} is not the same
as {\it statistical consistency}. In fact, a direct fit to the full
database with our model gives $\chi_{\rm min}^2/\nu=1.41$ which is
25$\sigma$ away from the expected value. This clearly indicates either
a bad model, inconsistent data, or both. A statistical measure of the
probability that the theory is plausible is given by the $p$-value;
assuming that the normality of residuals is correct it corresponds to
the probability of obtaining results at least as extreme as the
results actually observed~\cite{evans2004probability,
  eadie2006statistical}. Thus, the probability of having $\chi_{\rm
  min}^2/\nu=1.41$ for $\nu \sim 7000$ is $p=10^{-20}$, which clearly
rules out that the theory describes the data within fluctuations.

\subsection{Inconsistent vs consistent data}

The determination of theoretical uncertainties requires as a prerequisite the
compatibility or consistency of all data. This is a strong condition which is
not always fulfilled, particularly when the number of data becomes large. Most
often, different experiments have different sources of errors and are mutually
incompatible. Thus, while any statistical analysis benefits from a large
amount of data, a side effect is the proliferation of inconsistent data. In
that case it is obvious that no model will be able to simultaneously describe
all the data in a satisfactory manner. To appreciate this point more clearly,
assume two experiments which yield the measurements $O_{\rm exp1} \pm \Delta
O_{\rm exp1}$ and $O_{\rm exp2} \pm \Delta O_{\rm exp2}$. If the theoretical
estimate is $O_{\rm th}$, we have
\begin{equation}
  \chi^2 = \left[\frac{O_{\rm exp1}-O_{\rm th}}{\Delta O_{\rm exp1}} \right]^2 +
  \left[\frac{O_{\rm exp2}-O_{\rm th}}{\Delta O_{\rm exp2}} \right]^2 
\end{equation}
Minimizing respect to $O_{\rm th}$ we get
\begin{equation}
  \chi_{\rm min}^2 = \frac{(O_{\rm exp1}-O_{\rm exp2})^2}{\Delta
    O_{\rm exp1}^2+ \Delta O_{\rm exp2}^2}
\end{equation}
which becomes larger than 1 for $|O_{\rm exp1}-O_{\rm exp2}| \ge
\sqrt{\Delta O_{\rm exp1}^2+ \Delta O_{\rm exp2}^2}$, in which case we
have two {\it inconsistent} measurements. The important question is
whether both measurements are wrong or just only one. The term wrong
here does not necessarily mean an incorrect measurement; it suffices
if {\it one} or {\it both} errors $\Delta O_{\rm exp1}$ and $\Delta
O_{\rm exp2}$ are unrealistically small. In case of a discrepancy one
may re-analyze the experiment or simply ask the experts, an unfeasible
strategy for the experiments performed in the time span 1950-2013
comprising the analysis. The advantage of the statistical method is
that, for a large number of experiments, the systematic errors are
also randomized and one may rule out some experiments in a kind of
majority vote argument.

The case discussed previously corresponds to two different
measurements of the {\it same} observable, say the differential cross
section at the same energy and angle, and the generalization to any
number of experiments is straightforward. However in the case of
experiments with close kinematics there is no simple way to decide
between inconsistent data unless some continuity and smooth behavior
is assumed in order to intertwine the two measurements. Here is where
the {\it model} enters and statistical methods will never tell us if a
given model is correct but rather if the model is inconsistent with
the data. This is a kind of circular argument which can only be
avoided by looking for models which congregate as many data as
possible in a consistent way. Clearly, following this criterion, once
one finds a good model, any improvement of the model should describe
{\it more data} in a statistically significant fashion. The great
advantage is that if there are reasons to intertwine theoretically the
different measurements of all possible observables one may discuss the
data consistency in a generalized way and be able to select between
different observables.

\subsection{Self-consistent data selection}

The self-consistent criterion for data selection was proposed by Gross and
Stadler~\cite{Gross:2008ps} and implemented in \cite{Perez:2013mwa}. The way
data have been selected proceeds according to the following procedure:
\begin{enumerate}
\item Fit the model to all data.  If $\chi^2/\nu < 1$ you can stop. If not
  proceed further.
\item  Remove data sets with improbably high or low $\chi^2$
(3$\sigma$ criterion)
\item  Refit parameters for the remaining data. 
\item   Re-apply $3\sigma$ criterion to all data
\item   Repeat until no more data are excluded or recovered
\end{enumerate}
The effect of the selection criterion with our model is to go from
$\chi^2/\nu|_{\rm all}=1.41 $ to $ \chi^2/\nu|_{\rm selected}=1.05$ with a
reduction in the number of data from $N_{\rm Data}= 8173 $ to $ N_{\rm Data}=
6713$.  While this seems a drastic rejection it is the largest self-consistent
fit to date below 350 MeV. For this number of data this is {\it not} a minor
improvement; in terms of a normality test, it makes the difference in
$p$-value between having $p=10^{-20}$ or $p=0.68$.

\subsection{Fitting results}

The set of 32 scattering observables which we use for the fits comprises a total
of about 7000 selected measurements. It is interesting to decompose the
contributions to the total $\chi^2$ both in terms of the fitted observables as
well as in different energy bins. The separation is carried out explicitly in
Tables~\ref{tab:chi2-obs-pp} and \ref{tab:chi2-obs-np} for pp and np
scattering observables respectively and for the latest fit which includes also
the pion-nucleon coupling constants\cite{Perez:2016aol, Arriola:2016hfi} (see
below). As we can see the size of the contributions $\chi^2/N$ are at similar
levels for most observables. Note that observables with a considerable larger
or smaller $\chi^2/N$ are also observables with a small number of data and
therefore larger statistical fluctuations are expected (we remind that
for N {\it independent data} we expect $\chi^2/N \approx 1 \pm \sqrt{2/N}$.

\begin{table}[ht]
  \caption{\label{tab:chi2-obs-pp} Contributions to the total $\chi^2$
    for different pp
    observables\cite{Perez:2016aol,Arriola:2016hfi}. We use the
    notation of \cite{Hoshizaki:1969qt,Bystricky:1976jr}. } \small
  \begin{tabular*}{\columnwidth}{@{\extracolsep{\fill}}ccccccccccc}
    \hline
    Observable  & \phantom{11}Code\phantom{11} & $N_{pp}$ & $\chi_{pp}^2$ & $\chi_{pp}^2/N_{pp}$  \\     
    $d\sigma/d\Omega$ & DSG    &   935 &  903.5  &   0.97 \\
    $A_{yy}$          & AYY    &   312 &  339.0  &   1.09 \\
    $D$               & D      &   104 &  135.1  &   1.30 \\
    $P$               & P      &   807 &  832.4  &   1.03 \\
    $A_{zz}$          & AZZ    &    51 &   47.4  &   0.93 \\
    $R$               & R      &   110 &  112.8  &   1.03 \\
    $A$               & A      &    79 &   70.5  &   0.89 \\
    $A_{xx} $         & AXX    &   271 &  250.7  &   0.92 \\
    $C_{kp} $         & CKP    &     2 &    3.1  &   1.57 \\
    $R'$              & RP     &    29 &   11.9  &   0.41 \\
    $M_{s'0sn} $      & MSSN   &    18 &   13.1  &   0.73 \\
    $N_{s'0kn}$       & MSKN   &    18 &    8.5  &   0.47 \\
    $A_{zx}$          & AZX    &   264 &  250.6  &   0.95 \\
    $A'$              & AP     &     6 &    0.8  &   0.14 \\
    \botrule
  \end{tabular*}
\end{table}
\begin{table}[ht]
  \caption{\label{tab:chi2-obs-np} Contributions to the total $\chi^2$
    for different np
    observables\cite{Perez:2016aol,Arriola:2016hfi}. We use the
    notation of \cite{Hoshizaki:1969qt,Bystricky:1976jr} } \small
  \begin{tabular*}{\columnwidth}{@{\extracolsep{\fill}}ccccccccccc}
    \toprule
    Observable  & \phantom{11}Code\phantom{11} &  $N_{np}$ & $\chi_{np}^2$ & $\chi_{np}^2/N_{np}$  \\
    \hline 
    $d\sigma/d\Omega$ & DSG    &  1712   & 1803.4 & 1.05 \\
    $D_t$             & DT     &    88   &   83.7 & 0.95 \\
    $A_{yy}$          & AYY    &   119   &   96.0 & 0.81 \\
    $D$               & D      &    29   &   37.1 & 1.28 \\
    $P$               & P      &   977   &  941.7 & 0.96 \\
    $A_{zz}$          & AZZ    &    89   &  108.1 & 1.21 \\
    $R$               & R      &     5   &    4.5 & 0.91 \\
    $R_t$             & RT     &    76   &   72.2 & 0.95 \\
    $R_t'$            & RPT    &     4   &    1.4 & 0.35 \\
    $A_t$             & AT     &    75   &   77.0 & 1.03 \\
    $D_{0s''0k} $     & D0SK   &    29   &   44.0 & 1.52 \\
    $N_{0s''kn}$      & NSKN   &    29   &   25.5 & 0.88 \\
    $N_{0s''sn}$      & NSSN   &    30   &   20.3 & 0.68 \\
    $N_{0nkk}$        & NNKK   &    18   &   13.5 & 0.75 \\
    $A$               & A      &     6   &    2.9 & 0.49 \\
    $\sigma $         & SGT    &   411   &  500.2 & 1.22 \\
    $\Delta \sigma_T $& SGTT   &    20   &   26.3 & 1.31 \\
    $\Delta \sigma_L $& SGTL   &    16   &   18.4 & 1.15 \\
    \botrule
  \end{tabular*}
\end{table}

Likewise, we can also break up the contributions in order to see the
significance of different energy intervals, see Table~\ref{tab:bins}. We find
that, in agreement with the Nijmegen analysis (see \cite{Stoks:1993zz,
Stoks:1994pi} for comparisons with previous potentials), there is a relatively
large degree of uniformity in describing data at different energy bins. We
note also that the fit in the low energy region below $2$ MeV gives the
largest values for $\chi^2/N$. 

\begin{table*}[ht]
  \caption{\label{tab:bins} The $\chi^{2}$ results of the main
    combined $pp$ and $np$ partial-wave analysis
    \cite{Perez:2016aol,Arriola:2016hfi} for the 10 single-energy bins
    in the range $0 < T_{\rm LAB} < 350 {\rm MeV}$. We compare the fit
    $\chi^2/N|_{\rm fit}$ with the theoretical expectation
    $\chi^2/N|_{\rm th}=1\pm \sqrt{2/N}$.}  \small
  \begin{tabular*}{\textwidth}{@{\extracolsep{\fill}}ccccccccccc}
    \toprule
    Bin (MeV)  & $N_{pp}$ & $\chi_{pp}^2$ & $\chi_{pp}^2/N_{pp}$ & $N_{np}$ & $\chi_{np}^2$ & $\chi_{np}^2/N_{np}$ & $N $ & $\chi^2$ & $\chi^2/N|_{\rm fit}$ &$\chi^2/N|_{\rm th}$ \\
    \hline 
    0.0-0.5  & 103 &  107.2  & 1.04 &  46  &  88.2 & 1.92 &  149 &  195.4 & 1.31 & $1\pm$ 0.11 \\
    0.5-2    &  82 &   58.8  & 0.72 &  50  &  92.8 & 1.86 &  132 &  151.5 & 1.15 & $1\pm$ 0.12 \\
    2-8      &  92 &   80.1  & 0.87 & 122  & 151.0 & 1.24 &  214 &  231.0 & 1.08 & $1\pm$ 0.10\\ 
    8-17     & 124 &  100.3  & 0.81 & 229  & 183.9 & 0.80 &  353 &  284.1 & 0.80 & $1\pm$ 0.08\\
    17-35    & 111 &   85.5  & 0.77 & 346  & 324.2 & 0.94 &  457 &  409.7 & 0.90 & $1\pm$ 0.07\\
    35-75    & 261 &  231.2  & 0.89 & 513  & 559.7 & 1.09 &  774 &  790.9 & 1.02 & $1\pm$ 0.05\\
    75-125   & 152 &  154.8  & 1.02 & 399  & 445.2 & 1.12 &  551 &  600.0 & 1.09 & $1\pm$ 0.06\\
    125-183  & 301 &  300.5  & 1.00 & 372  & 381.7 & 1.03 &  673 &  682.2 & 1.01 & $1\pm$ 0.05\\
    183-290  & 882 &  905.0  & 1.03 & 858  & 841.4 & 0.98 & 1740 & 1746.4 & 1.00 & $1\pm$ 0.03\\
    290-350  & 898 &  956.1  & 1.06 & 798  & 808.1 & 1.01 & 1696 & 1764.1 & 1.04 & $1\pm$ 0.03\\
    \botrule
  \end{tabular*}
\end{table*}

From the optimal fitting parameters $V^{\alpha} (r_n)$ with
$\alpha=^1S_0,^3P_0,^3S_1,^3D_1,E_1, \dots$ being the different partial waves
in a given pp or np channel, we define $(\lambda_n)^{\alpha} = 2\mu_{ab}
V^{\alpha} (r_n) \Delta r$ which has units of ${\rm fm}^{-1}$ and $ab=pp,np$.
In table \ref{tab:FitParameters} we show the corresponding numerical values.
It would be nice to see whether something can be said about the nn
interaction. However, one remarkable feature of this and similar analyses is
the fact that with the exception of S-waves the short distance parameters can
be chosen to coincide in the pp and np systems with common partial waves. The
fact that to this date it is not possible to do it for S-waves precludes to
{\it} predict the nn interaction from the combined $np$ and $pp$ fit (see
however a theoretical discussion in Ref.~\cite{CalleCordon:2010sq}).

{\small 
\begin{table}[tb]
  \caption{\label{tab:FitParameters} Fitting delta-shell parameters
    $(\lambda_n)^{JS}_{l,l'} $ (in ${\rm fm}^{-1}$) with their errors
    for all states in the $JS$ channel for a fit with isospin symmetry
    breaking on the $^1S_0$ partial wave parameters only and the
    pion-nucleon coupling constants $f^2_0$, $f^2_p$ and $f^2_c$ as
    fitting parameters We take $N=5$ equidistant points with $\Delta r
    = 0.6$fm. $-$ indicates that the corresponding fitting
    $(\lambda_n)^{JS}_{l,l'} =0$. The lowest part of the table shows
    the resulting OPE coupling constants with errors}
    \begin{tabular*}{\textwidth}{@{\extracolsep{\fill}}l D{.}{.}{2.6} D{.}{.}{2.6} D{.}{.}{2.6} D{.}{.}{2.6} D{.}{.}{2.6} }
      Wave  & \multicolumn{1}{c}{$\lambda_1$} & 
      \multicolumn{1}{c}{$\lambda_2$} & 
      \multicolumn{1}{c}{$\lambda_3$} & 
      \multicolumn{1}{c}{$\lambda_4$} & 
      \multicolumn{1}{c}{$\lambda_5$}  \\
      \hline\noalign{\smallskip}
      $^1S_{0{\rm np}}$& 1.16(6)               &-0.77(2)               &-0.15(1)               &\multicolumn{1}{c}{$-$}&-0.024(1)              \\
      $^1S_{0{\rm pp}}$& 1.31(2)               &-0.716(5)              &-0.192(2)              &\multicolumn{1}{c}{$-$}&-0.0205(4)             \\
      $^3P_0$          &\multicolumn{1}{c}{$-$}& 0.94(2)               &-0.319(7)              &-0.062(3)              &-0.023(1)              \\
      $^1P_1$          &\multicolumn{1}{c}{$-$}& 1.20(2)               &\multicolumn{1}{c}{$-$}& 0.075(2)              &\multicolumn{1}{c}{$-$}\\
      $^3P_1$          &\multicolumn{1}{c}{$-$}& 1.354(5)              &\multicolumn{1}{c}{$-$}& 0.0570(5)             &\multicolumn{1}{c}{$-$}\\
      $^3S_1$          & 1.79(7)               &-0.47(1)               &\multicolumn{1}{c}{$-$}&-0.072(2)              &\multicolumn{1}{c}{$-$}\\
      $\varepsilon_1$  &\multicolumn{1}{c}{$-$}&-1.65(2)               &-0.33(2)               &-0.233(7)              &-0.018(3)              \\
      $^3D_1$          &\multicolumn{1}{c}{$-$}&\multicolumn{1}{c}{$-$}& 0.40(1)               & 0.070(9)              & 0.021(3)              \\
      $^1D_2$          &\multicolumn{1}{c}{$-$}&-0.20(1)               &-0.206(3)              &\multicolumn{1}{c}{$-$}&-0.0187(3)             \\
      $^3D_2$          &\multicolumn{1}{c}{$-$}&-1.01(3)               &-0.17(2)               &-0.237(6)              &-0.016(2)              \\
      $^3P_2$          &\multicolumn{1}{c}{$-$}&-0.482(1)              &\multicolumn{1}{c}{$-$}&-0.0289(7)             &-0.0037(4)             \\
      $\varepsilon_2$  &\multicolumn{1}{c}{$-$}& 0.32(2)               & 0.190(4)              & 0.050(2)              & 0.0127(6)             \\
      $^3F_2$          &\multicolumn{1}{c}{$-$}& 3.50(6)               &-0.229(5)              &\multicolumn{1}{c}{$-$}&-0.0140(5)              \\
      $^1F_3$          &\multicolumn{1}{c}{$-$}&\multicolumn{1}{c}{$-$}& 0.12(2)               & 0.089(8)              &\multicolumn{1}{c}{$-$}\\
      $^3D_3$          &\multicolumn{1}{c}{$-$}& 0.54(2)               &\multicolumn{1}{c}{$-$}&\multicolumn{1}{c}{$-$}&\multicolumn{1}{c}{$-$}\\
      \hline\noalign{\smallskip}
      \multicolumn{2}{c}{$f^2_{\rm p}$} & \multicolumn{2}{c}{$f^2_0$} & \multicolumn{2}{c}{$f^2_{\rm c}$} \\
      \noalign{\smallskip}
      \hline\noalign{\smallskip}
      \multicolumn{2}{c}{0.0764(4)} & \multicolumn{2}{c}{0.0779(8)} & \multicolumn{2}{c}{0.0758(4)} \\
    \end{tabular*}
\end{table}
}

\subsection{Covariance matrix error analysis and statistical correlations}
\label{sec:corr}

After the data selection and fitting, error propagation becomes applicable. 
Here we show the results for the conventional covariance error analysis which
assumes small errors and where one first determines the uncertainty in the
fitting parameters $V_i(r_n)$ which will be labeled generically as $\lambda_i$
for ease of notation~\footnote{ The bootstrap approach based on the MonteCarlo
method~\cite{Nieves:1999zb, Perez:2013mwa} will be discussed below.}.

Expanding around the minimum values, $\bar \lambda_i$ has
\begin{equation}
\chi^2 = \chi^2_{\rm min}+ \sum_{ij=1}^{N_P} (\lambda_i-\bar \lambda_{i})
(\lambda_j-\bar \lambda_{j}) {\mathcal E}^{-1}_{ij} + \cdots
\end{equation}
where the $N_P\times N_P$ 
error matrix is defined as the inverse of the Hessian matrix
evaluated at the minimum
\begin{equation}
 {\mathcal E}^{-1}_{ij}= \frac12 
\frac{\partial^2 \chi^ 2}{\partial \lambda_i
  \partial \lambda_j}\Big|_{\lambda_i = \bar \lambda_i}
\end{equation}
The correlation matrix between the fitting parameters
$\lambda_i$ and $\lambda_j$ is given by
\begin{equation}
 {\mathcal C}_{ij}=  \frac{{\mathcal E}_{ij}}{\sqrt{{\mathcal E}_{ii}{\mathcal E}_{jj}}}
\label{eq:correlation}
\end{equation}
We compute the error of the parameter $\lambda_i$ as
\begin{equation}
\Delta\lambda_i \equiv \sqrt{{\mathcal E}_{ii}} .
\end{equation}
Error propagation of an observable depending on the fitting parameters
$G=G(\lambda_1,\ldots,\lambda_P)$ is computed as
\begin{equation}
(\Delta G)^2 = \sum_{ij} 
\frac{\partial G }{\partial\lambda_i}
\frac{\partial G}{\partial\lambda_j}  \Big|_{\lambda_k=\lambda_{k,0}}
{\mathcal E}_{ij}.
\label{eq:error-prop}
\end{equation}
The correlation matrix, Eq.~(\ref{eq:correlation}), has been evaluated in
Refs.~\cite{Perez:2014yla,Perez:2014kpa} where it has been found that for the
potentials {\it in the partial wave basis} $V_{l,l'}^{JS} (r_n)$ the different
points $r_n$ are largely correlated within a given partial wave, whereas {\it
different} partial waves are largely uncorrelated. This information allows to
substantially speed up the minimum search as movements in the multidimensional
space are thus independent and the approaching path to the minimum operates
stepwise~\cite{Perez:2014yla,Perez:2014kpa}.

\subsection{Phase-shifts}
\label{sec:phases} 

The first useful application of error propagation regards scattering
amplitudes and phase shifts. Extensive tables for the selected values $T_{\rm
LAB}=1,5,10,25,50,100,150,200,250,300,350$ MeV have traditionally been
presented since the Nijmegen analysis as representative of the fits. These
energy values corresponds to a grid of almost equidistant  CM momenta $p=
\sqrt{T_{\rm LAB} M_N/2}$ between 0 and 2   fm$^{-1}$.

\begin{figure}[tb]
\begin{center}
  \epsfig{figure=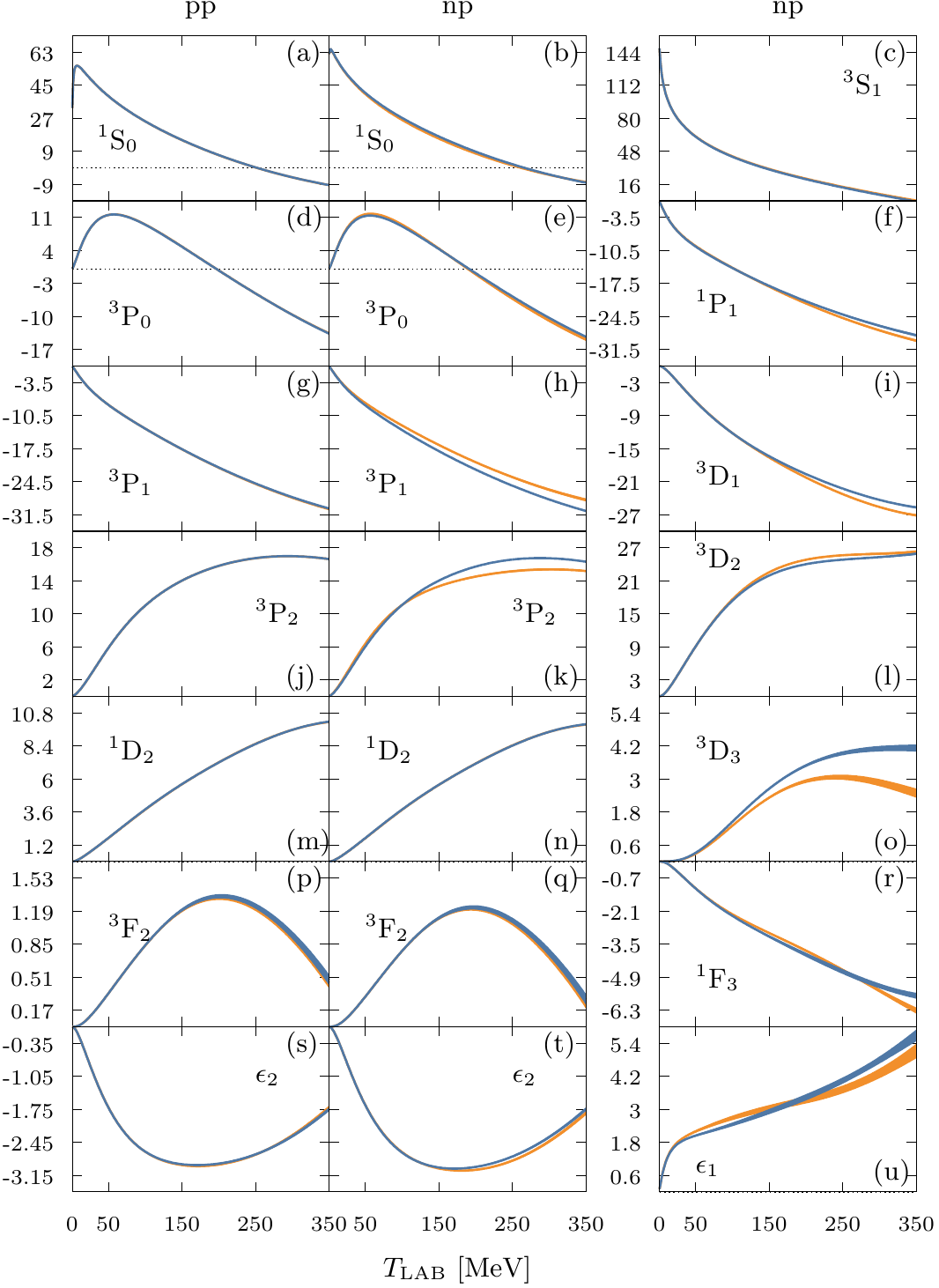,width=.7\linewidth} 
\end{center}
\caption{(Color online) Phase shifts obtained from a partial waves
  analysis to pp and np data and statistical uncertainties. Blue band
  from~\cite{Perez:2013jpa} from a fit with {\it fixed} $f^2$ 
  and red band~~\cite{Perez:2016aol} from a fit with charge
  symmetry breaking on the $^3P_0$, $^3P_1$ and $^3P_2$ partial
  waves {\it and} in the coupling constants $f_0^2$, $f_p^2$ and $f_c^2$.}
\label{Fig:Phaseshifts}
\end{figure}

For illustration, Fig.~\ref{Fig:Phaseshifts} compares, for low angular
momentum, the phase shifts of the primary PWA in~\cite{Perez:2013jpa}
from a fit with {\it fixed} pion coupling constant, $f^2$, (blue
bands) and the most recent ones~\cite{Perez:2016aol} (red band) from a
fit with charge symmetry breaking on the $^3P_0$, $^3P_1$ and $^3P_2$
partial waves {\it and} in the pion coupling constants $f_0^2$,
$f_p^2$ and $f_c^2$.

\section{Determination of Yukawa coupling constants}
\label{sec:yukawa}

The first determination of the coupling constant was carried out in
1940 by Bethe who obtained the value $f^2=0.077-0.080$ from the study
of deuteron properties~\cite{Bethe:1940zz} and very close to the
currently accepted value (see table~\ref{tab:Constants}). Subsequent
determinations based on a variety of processes can be traced from
recent compilations~\cite{deSwart:1997ep, Sainio:1999ba}. A recent
historical account has been given by Matsinos~\cite{Matsinos:2019kqi}
where some newer determinations can be consulted according to his own
eligibility criterium. For completeness we also quote recent studies
based on pion-deuteron scattering~\cite{Baru:2010xn,Baru:2011bw} or on
the analysis of Roy equations for $\pi N$~\cite{Hoferichter:2015hva}
where an upgrade of the corresponding scattering data is considered.

We note that what follows is a brief summary of the results presented
in our previous papers where many more details may be found regarding
the most influential observables, the dependence on the cut-off radius
$r_c$, the inclusion of two-pion exchange contributions or the energy
range used in the fit or the evolution with the numerical values and
precision along the years~\cite{Perez:2016aol,Arriola:2016hfi}.

\begin{table*}[tb]
  \caption{\label{tab:fpf0fc} The pion-nucleon coupling constants $f_p^
    2$, $f_0^ 2$ and $f_c^ 2$ determined from different fits to the
    Granada-2013 database and their characteristics. We indicate the
    partial waves where charge dependence is allowed.}
  \begin{tabular*}{\textwidth}{@{\extracolsep{\fill}}cccccccccc}
    \toprule
    $f_p^2$ & $f_0^2$ & $f_c^2$ & CD-waves & $\chi^2_{pp}$ &  $\chi^2_{np}$ & $\chi^2$ & $N_{\rm Dat}$ & $N_{\rm Par}$ & $\chi^2/\nu $ \\ 
    \hline
    0.075     & idem      & idem      & $^1S_0$      & 2997.29 & 3957.57 & 6954.86 & 6720 & 46 & 1.042 \\ 
    0.0763(1) & idem      & idem      & $^1S_0$      & 2995.20 & 3952.85 & 6947.05 & 6720 & 47 & 1.041 \\ 
    0.0764(4) & 0.0779(8) & 0.0758(4) & $^1S_0$      & 2994.41 & 3950.42 & 6944.83 & 6720 & 49 & 1.041 \\
    0.0761(4) & 0.0790(9) & 0.0772(5) & $^1S_0$, $P$ & 2979.37 & 3876.13 & 6855.50 & 6741 & 55 & 1.025 \\  
    \botrule
  \end{tabular*}
\end{table*}

The $\pi NN$ coupling constant is defined as the pion-nucleon-nucleon vertex
when the three particles are on the mass shell. The corresponding potentials
would be 
\begin{align}
 V_{pp \to pp}(r) =& f^2_{\pi^0 pp} V_{m_{\pi^0}}(r),  \\
 V_{np \to np}(r) = V_{pn \to pn}(r) 
=& -f_{\pi^0 nn}f_{\pi^0 pp}V_{m_{\pi^0}}(r) \\ 
V_{pn \to np}(r) = V_{np \to pn}(r) =& f_{\pi^-pn} f_{\pi^+np} \, 
 V_{m_{\pi^\pm}}(r) \\ 
V_{nn \to nn}(r) =& f^2_{\pi^0 nn} V_{m_{\pi^0}}(r),  
 \label{eq:BreakIsospinOPE2}
\end{align}
There exist four pion nucleon coupling constants, $f_{\pi^0 pp}$, $-f_{\pi^0
nn}$, $f_{\pi^+ pn} /\sqrt{2}$ and $ f_{\pi^- np} /\sqrt{2}$ which coincide
with $f$ when up and down quark masses are identical and the electron charge
is zero. In NN interactions we have access to the combinations, 
\begin{equation}
f_n^2 = f_{\pi^0 nn} f_{\pi^0 nn} \, , \quad   
f_p^2 = f_{\pi^0 pp} f_{\pi^0 pp} \, , \quad 
f_0^2 = -f_{\pi^0 nn} f_{\pi^0 pp} \, , \quad 
2 f_c^2 = f_{\pi^-pn} f_{\pi^+np} \, . 
\end{equation}

While there is no reason why the pion-nucleon-nucleon coupling constants
should be identical in the real world, one expects that the small differences
might be pinned down from a sufficiently large number of independent and
mutually consistent data. Note that from np and pp analysis we would obtain
$f_p^2$, $f_0^2$ and $f_c^2$ we may deduce the nn coupling using the previous
equations $f_n = - f_0^2/ f_p $.  We try to find out how many data would be
needed by recalling that electroweak corrections scale with the fine structure
constant $\alpha=1/137$ and the light quark mass differences. Thus
\begin{equation}
\frac{\delta g}{g} = {\mathcal O} \left(\alpha, \frac{m_u-m_d}{\Lambda_{\rm QCD}} \right) = {\mathcal O} \left(\alpha, \frac{M_p-M_n}{\Lambda_{\rm QCD}} \right) 
\end{equation}
for the relative change around a mean value. These are naturally at the
$1-2\%$ level, a small effect. The question is on how many independent
measurements $N$ are needed to achieve this desired accuracy. According to the
central limit theorem, for $N$ direct independent measurements the relative
standard deviation scales as
$$
\frac{\Delta g}{g} = {\mathcal O} \left( \frac{1}{\sqrt{N}} 
\right) 
$$
and $\delta g \sim \Delta g$ for $N=7000-10000$. We cannot carry out these
direct measurements of $g$ but we can proceed indirectly by considering a set
of mutually consistent NN scattering measurements The most recent analysis
\cite{Perez:2016aol,Arriola:2016hfi} based on the Granada-2013 database
comprises 6713 published data. This allows: i) to reduce the error bars, as
expected and ii) to discriminate between the three coupling constants (see
Table~\ref{tab:fpf0fc}).  When charge dependence in $^1S_0$, $P$ waves is
allowed one has 
\begin{equation}
f_p^2 = 0.0761(4) \, , \quad f_{0}^2 = 0.0790(9) \,, \quad f_{c}^2 =
0.0772(5) \, ,
\end{equation}
The most remarkable consequence is that from the point of view of the strong
interaction neutrons interact more strongly than protons.

\section{Systematic vs statistical errors: The 6 Granada potentials}
\label{sec:6Gr}

Within the phenomenological approach the estimation of systematic errors can
be addressed by using different representations of the mid-range function
below the separation distance $r_c$ while keeping the long range potential and
the NN database. To this end we have analyzed 6 different potentials in
Ref.~\cite{Perez:2014waa} which have been fitted to the {\it same} Granada
2013 database and have the {\it same} long distance components of the
potential. First we have checked that the  6 Granada potentials are
statistically acceptable. In fact, as it has been stressed in our previous
works~\cite{Perez:2014yla, Perez:2014kpa} one can globally slightly enlarge
the experimental uncertainties by the so-called Birge
factor~\cite{birge1932calculation} provided the residuals verify a normality
test. After this re-scaling the $p$-value becomes 0.68 for a $1\sigma$
confidence level and hence all potentials become statistically equivalent. The
results are summarized in Table~\ref{tab:PotentialsSummary}. Thus, the overall
spread between the various phenomenological models with $\chi^2 / {\rm dof}
\sim 1$ provides an estimate of the scale of the systematic uncertainty.  A
direct way of illustrating quantitatively the situation is by analyzing the
corresponding phase shifts in the different analyses.

\begin{table*}[ht]
 \caption{\label{tab:PotentialsSummary} Granada  Potentials Summary.}
 \begin{tabular*}{\textwidth}{@{\extracolsep{\fill}} l *9{c}}
   Potential       & $N_{\rm Par}$ & $N_{np}$ & $N_{pp}$ & $\chi^2_{np}$ & $\chi^2_{pp}$ & $\chi^2/{\rm d.o.f.}$ & p-value  & Normality & 
 Birge Factor \\
   \hline
   DS-OPE          & 46         & 2996     & 3717     & 3051.64       & 3958.08    & 1.05     & 0.32  &  Yes &  1.03   \\
   DS-$\chi$TPE    & 33         & 2996     & 3716     & 3177.43       & 4058.28    & 1.08     & 0.50  & Yes  &  1.04   \\
   DS-$\Delta$BO   & 31         & 3001     & 3718     & 3396.67       & 4076.43    & 1.12     & 0.24  & Yes  &  1.06   \\
   Gauss-OPE       & 42         & 2995     & 3717     & 3115.16       & 4048.35    & 1.07     & 0.33  & Yes  &  1.04   \\
   Gauss-$\chi$TPE & 31         & 2995     & 3717     & 3177.22       & 4135.02    & 1.09     & 0.23  & Yes  &  1.05  \\
   Gauss-$\Delta$BO& 30         & 2995     & 3717     & 3349.89       & 4277.58    & 1.14     & 0.20  & Yes  &  1.07     
 \end{tabular*}
\label{tab:gr-pots}
\end{table*}

\begin{figure}[ht]
\begin{center}
\epsfig{figure=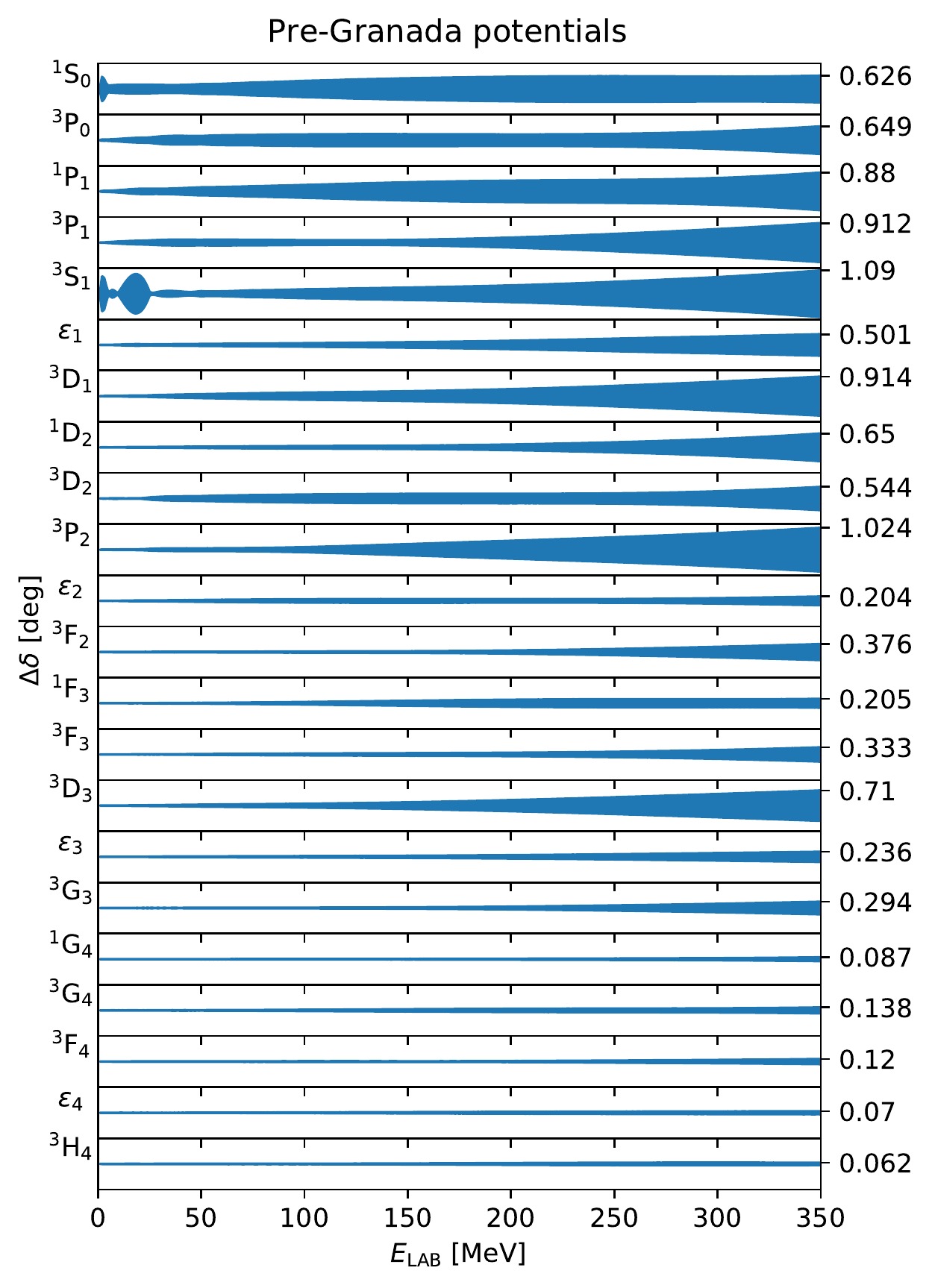,width=6.5cm}
\hskip.5cm
\epsfig{figure=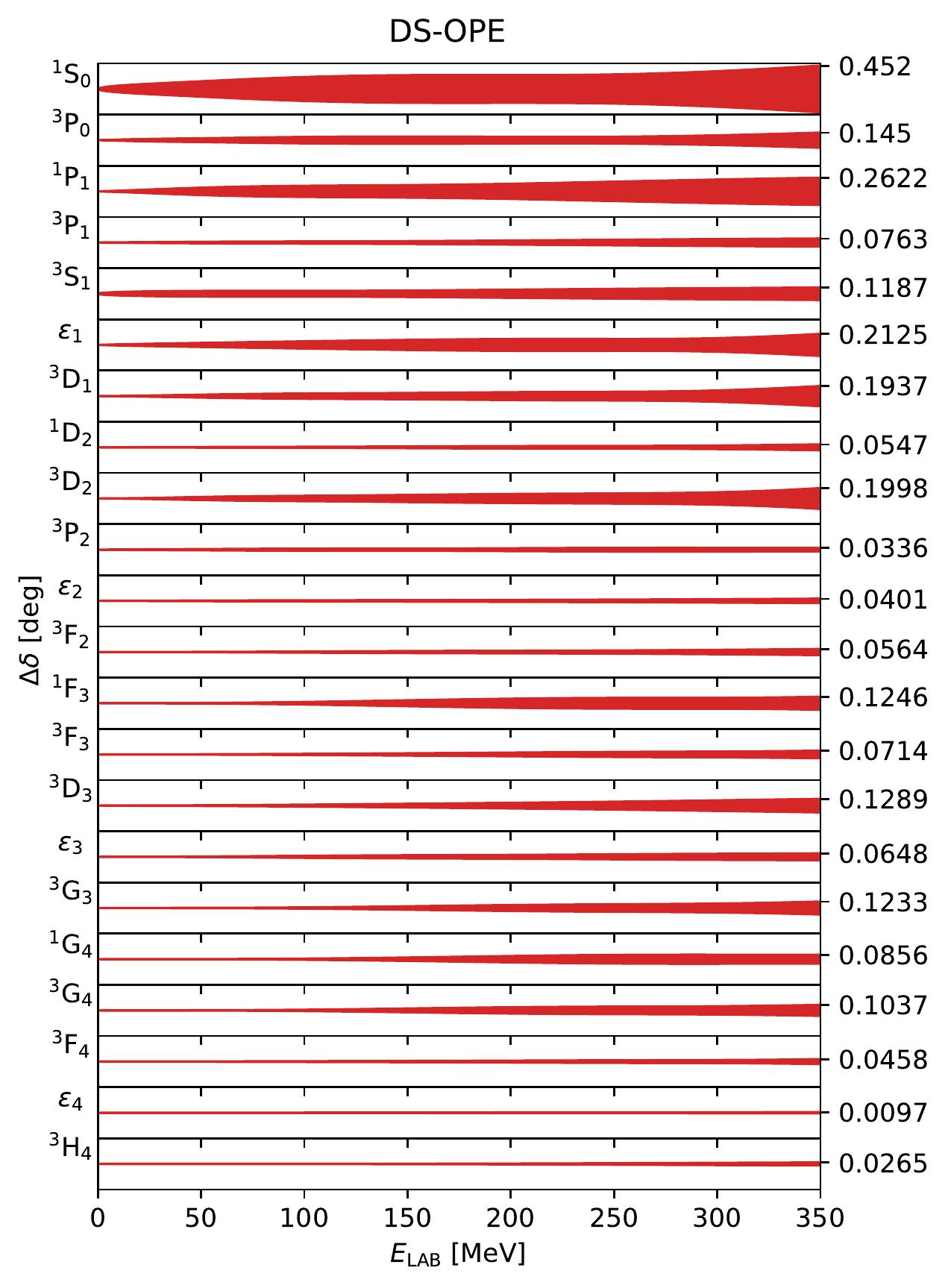,width=6.5cm}
\hskip.5cm
\epsfig{figure=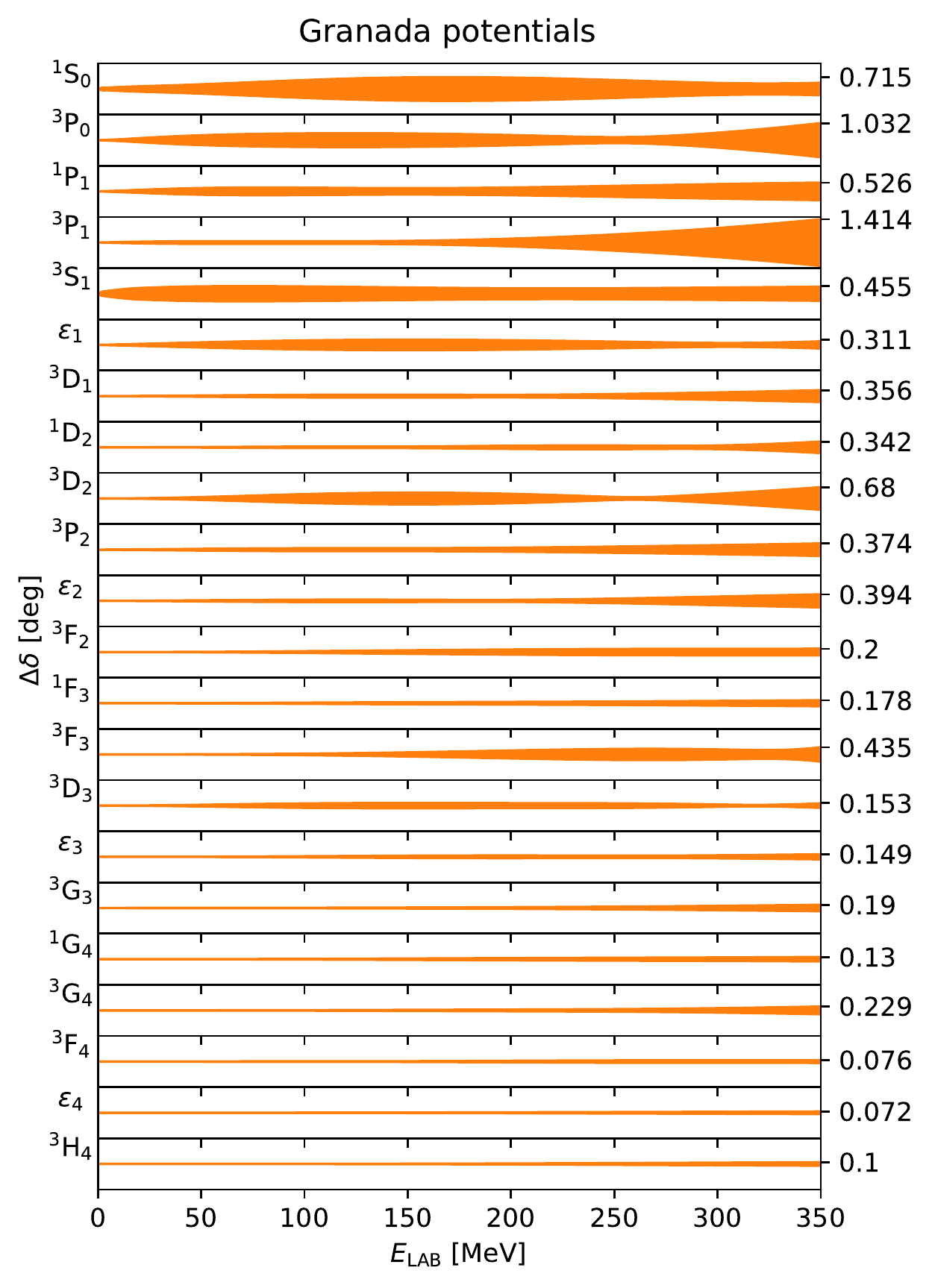,width=6.5cm}
\hskip.5cm
\epsfig{figure=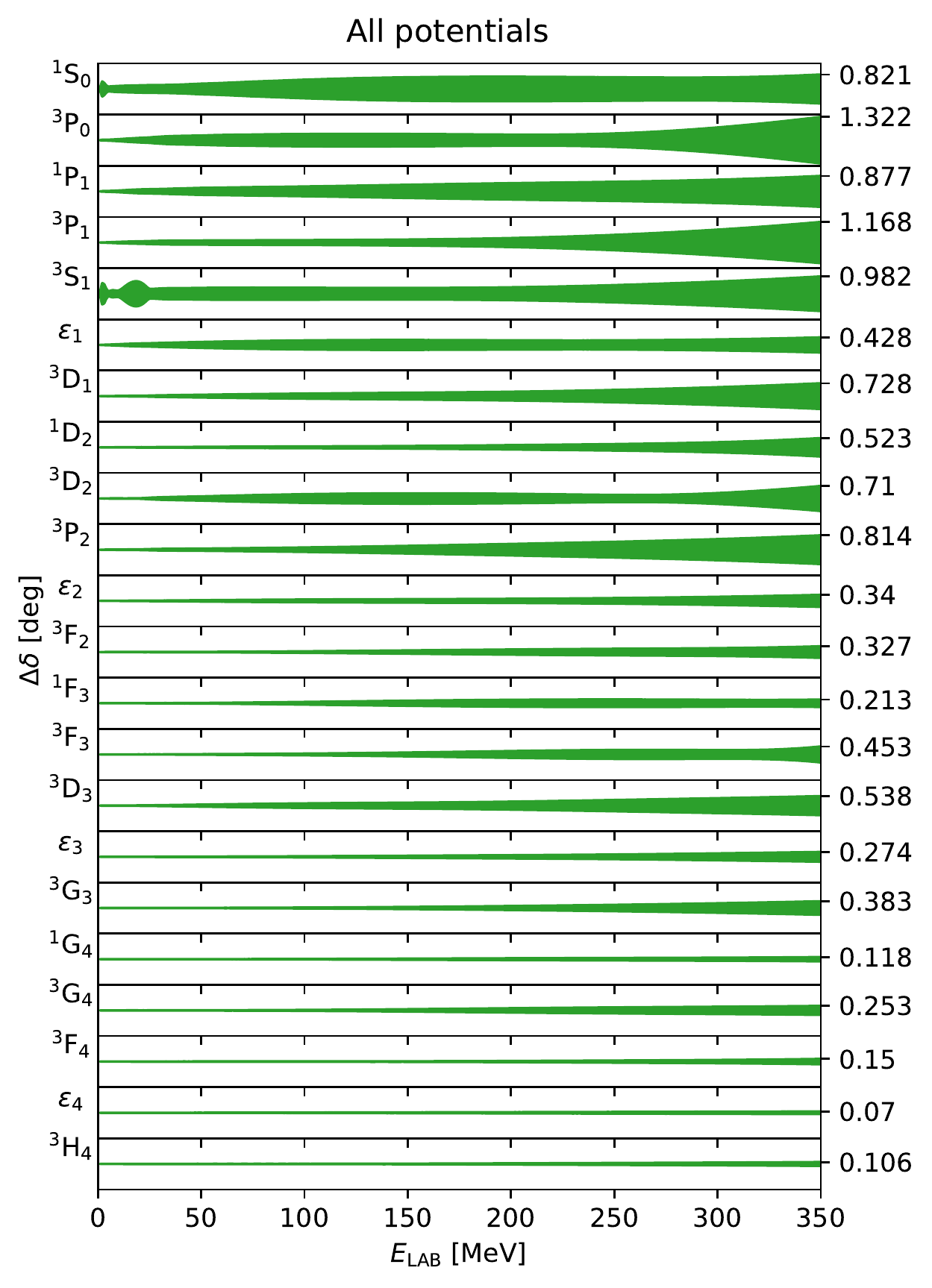,width=6.5cm}
\end{center}
\caption{ Uncertainties (in degrees, right axis) for partial wave np
  phase shifts with $J \le 4$ (left axis) for different potentials
  fitting scattering data with $\chi^2/{\rm dof} \sim 1$ as a function
  of the LAB energy (in MeV). (Upper left panel) Averaged errors for
  pre-Granada
  potentials~\cite{Stoks:1993tb,Stoks:1994wp,Wiringa:1994wb,Machleidt:2000ge,Gross:2008ps}.
  (Upper right panel) statistical errors for the primary Granada 2013
  $\chi^2$ analysis~\cite{Perez:2013mwa}. (Lower left panel) the
  averaged errors for the 6 Granada
  potentials~\cite{Perez:2013mwa,Perez:2013oba,Perez:2014waa}. (Lower
  right panel) Averaged errors for all 13=7 pre-Granada and the 6
  Granada potentials.}
\label{fig:errors-ps}
\end{figure}

Thus, for each energy and partial wave, one evaluates the phaseshifts
$\delta^{(1)},\ldots,\delta^{(N)}$ for a representative set of high-precision
NN potentials $V^{(1)},\ldots,V^{(N)}$, and computes the average
$\overline{\delta}$ and standard deviation
\begin{equation}
\Delta\delta = 
\sqrt{
\frac{1}{N-1}
\sum_{i=1}^N 
\left(\delta^{(i)}-\overline{\delta}\right)^2}
\end{equation} 
as a measure of the systematic uncertainty of the  phaseshifts. In
Fig.~\ref{fig:errors-ps} we show the results for four different situations. To
provide some historical perspective, we show in the upper left panel the
averaged phase shifts, i.e. the absolute (mean-square) errors for np partial
wave phase shifts due to the different potentials fitting scattering data with
$\chi^2/{\rm dof} \sim 1$ \cite{Stoks:1993tb, Stoks:1994wp, Wiringa:1994wb,
Machleidt:2000ge, Gross:2008ps} as a function of the LAB energy, namely (CD
Bonn) \cite{Machleidt:1995km}, Nijmegen (Nijm-I and Nijm-II)
\cite{Stoks:1993tb}, Argonne AV18 \cite{Wiringa:1994wb}, Reid (Reid93)
~\cite{Friar:1993kk} and the covariant spectator model \cite{Gross:2008ps}. 
As one naturally expects the average uncertainties grow with energy and
decrease with the relative angular momentum which semi-classically corresponds
to probing an impact parameter 
\begin{equation} \label{impact}
 b =  \frac{L+1/2}{ p} 
\end{equation}
where $p$ is the CM momentum, $ p= \sqrt{M_N E_{\rm LAB} /2} $, making
peripheral waves to be mostly determined from OPE. These analyses stop
at the pion production threshold so that one probes distances larger
than
\begin{equation}
b_{\rm min} \sim 1/\Lambda = 0.5 {\rm fm}, \kern 1cm
\Lambda=\sqrt{m_\pi M_N}.
\end{equation}
Note that the bumps or bulges at low energy in $^1S_0$ and $^3S_1$ channels in
the top left panel are due to a unique potential which is an outlier at low
energies. In particular, the authors believe that the outlier behavior is due
to the use of an interpolating function used to approximate the potential
between the values of laboratory energy at which phaseshifts are usually
tabulated.

In the upper right panel of Fig.~\ref{fig:errors-ps} we show the
errors obtained via the standard covariance-matrix method explained
above and including correlations in the fitting parameters for the
primary Granada 2013 analysis~\cite{Perez:2013mwa} which corresponds
to the DS-OPE potential. Thirdly, in the lower left panel we show the
case of the np phase shifts for the 6 Granada
potentials~\cite{Perez:2013mwa, Perez:2013oba,
  Perez:2014waa}. Finally, in lower right panel we present the
uncertainties for all the 7 pre-Granada potentials and the 6 Granada
potentials simultaneously.

In Ref.~\cite{Perez:2014waa} we found similar statistical errors in  all the
Granada potentials, which are statistically validated with the {\it same}
Granada-2013 database, i.e., if the phase-shift for potential $V^{(i)}$ in a
given partial wave is $\delta^{(i)} \pm \Delta \delta_{\rm stat}^{(i)} $, then
\begin{equation}
 \Delta \delta_{\rm stat}^{(1)} \sim \dots \sim \Delta \delta_{\rm
   stat}^{(6)} \, ,
\end{equation}
However we also found that the standard deviation of systematic errors  obeys
\begin{equation}
\Delta \delta_{\rm sys} \equiv {\rm Std} (\delta^{(1)}, \dots ,
\delta^{(6)}) \gg \Delta \delta_{\rm stat}^{(i)} \, .
\end{equation}
In all the potentials, the tails  {\it above} $r=3$ fm (including CD-OPE and
all electromagnetic effects) are the {\it same}, thus the discrepancies
between the potentials at short distances dominate the uncertainties, rather
than the np and pp experimental data themselves.  This conclusion holds also
when all high quality potentials are considered~\cite{Perez:2014waa}. This
counter-intuitive result relies not only on the specific forms of potentials
which treat the mid-- and short-range behavior of the interaction differently
but also on the fact that the fits are mainly done to scattering amplitudes
rather than to the phase-shifts themselves.

\section{Low energy behavior}

\subsection{Low energy parameters}
\label{sec:lep}

The effective range expansion was proposed by
Bethe~\cite{Bethe:1949yr} in order to provide a model independent
characterization of the scattering at low energies where the shape of the potential is largely irrelevant. The extension to
higher partial waves reads (see e.g.  \cite{madsen2002effective})
\begin{equation}
k^{2l+1}  M_l(k) \equiv k^{2l+1} \cot{\delta_l(k)} = -\frac{1}{\alpha_l} + \frac12 r_l k^2 + v_{2,l} k^4 + v_{3,l} k^6 + \cdots
\end{equation}
where $\alpha_l$ is the scattering length, $r_l$ the effective range and
$v_{i,l}$ the curvature parameters. In the case of coupled channels due to the
tensor force one has that ${\bf S}^{JS} = ({\bf M}^{JS} - i {\bf 1})({\bf
M}^{JS} + i {\bf 1})^{-1} $ with $({\bf M}^{JS})^\dagger = {\bf M}^{JS}$ a
hermitian coupled channel matrix (also known as the K-matrix). At the level of
partial waves the multi-pion exchange diagrams generate left hand cuts in the
complex s-plane, which arise in addition to the NN elastic right cut and the
$\pi NN$, $ 2\pi NN$ etc., pion production cuts. At low energies for $ |p| \le
m_\pi/2$ we have~\cite{PavonValderrama:2005ku}
\begin{equation}
 p^{l+l'+1} M_{l,l'}^{JS} (p) = -(\alpha^{-1})^{JS}_{l,l'} + \frac12  (r)^{JS}_{l,l'}p^2+  (v)^{JS}_{l,l'}p^4 + \dots
\label{eq:ERE-coup}
\end{equation}
which is the coupled channels effective range expansion. While at lowest
orders explicit formulas where available in terms of wave functions, larger
order and partial waves become rather cumbersome and no practical formula
exists.

Fortunately, the variable S-matrix approach of
Calogero~\cite{calogero1967variable} offers a unique way to extract
low-energy threshold parameters for a given NN potential which was
extended to coupled channels~\cite{PavonValderrama:2005ku} and applied
to the Reid93 and NijmII potentials up to $J \le 5$. For the 6 Granada
potentials these have also been extracted and we have found that the
systematic uncertainties are generally at least an order of magnitude
larger than statistical uncertainties~\cite{Perez:2014waa}. In
table~\ref{tab:LEPS-statistic} where we provide the low energy
parameters for $J \le 2$).

\def\EP{\epsilon}
\begin{table}
 \caption{\label{tab:LEPS-statistic} Low energy threshold np
   parameters for all partial waves with $j \leq 2$. The central value
   and \emph{statistical} error bars are given on the first line of
   each partial wave and correspond to the mean and standard deviation
   of a population of $1020$ parameters calculated with the Monte
   Carlo family of potential parameters described
   in~\cite{Perez:2014jsa} using the DS-OPE
   potential~\cite{Perez:2013mwa,Perez:2013jpa}.  The second line
   quotes the \emph{systematic} uncertainties, the central value and
   error bars correspond to the mean and standard deviation of the 9
   realistic potentials NijmII~\cite{Stoks:1994wp},
   Reid93~\cite{Stoks:1994wp}, AV18~\cite{Wiringa:1994wb},
   DS-OPE~\cite{Perez:2013mwa,Perez:2013jpa},
   DS-$\chi$TPE~\cite{Perez:2013oba,Perez:2013cza},
   Gauss-OPE~\cite{Perez:2014yla}, Gauss-$\chi$TPE, DS-$\Delta$BO and
   Gauss-$\Delta$BO. For each partial wave we show the scattering
   length $\alpha$ and the effective range $r_0$, both in ${\rm
     fm}^{l+l'+1}$, as well as the curvature parameters $v_2$ in ${\rm
     fm}^{l+l'+3}$, $v_3$ in ${\rm fm}^{l+l'+5}$ and $v_4$ in ${\rm
     fm}^{l+l'+5}$. For the coupled channels we use the nuclear bar
   representation of the $S$ matrix. Uncertainties smaller than
   $10^{-3}$ are not quoted} {\small 
  \begin{tabular*}{\columnwidth}{@{\extracolsep{\fill}} l D{.}{.}{3.5}
 D{.}{.}{3.5} D{.}{.}{3.5} D{.}{.}{3.5} D{.}{.}{4.7} }
 Wave   & \multicolumn{1}{c}{$\alpha$} 
        & \multicolumn{1}{c}{$ r_0$} 
        & \multicolumn{1}{c}{$ v_2$} 
        & \multicolumn{1}{c}{$ v_3$} 
        & \multicolumn{1}{c}{$ v_4$} \\
 \hline \noalign{\smallskip}
$^1S_0$ &   -23.735(6)  &    2.673(9)  &    -0.50(1)  &     3.87(2)  &   -19.6(1)   \\
        &   -23.735(16) &    2.68(3)   &    -0.48(2)  &     3.9(1)   &   -19.6(5)   \\
$^3P_0$ &    -2.531(6)  &    3.71(2)   &     0.93(1)  &     3.99(3)  &    -8.11(5)  \\
        &    -2.5(1)    &    3.7(4)    &     0.9(5)   &     3.9(1)   &    -8.2(9)   \\
$^1P_1$ &     2.759(6)  &   -6.54(2)   &    -1.84(5)  &     0.41(2)  &     8.39(9)  \\
        &     2.78(3)   &   -6.46(9)   &    -1.7(2)   &     0.5(2)   &     8.0(3)   \\
$^3P_1$ &     1.536(1)  &   -8.50(1)   &     0.02(1)  &    -1.05(2)  &     0.56(1)  \\
        &     1.52(1)   &   -8.6(1)    &    -0.06(7)  &    -0.9(2)   &     0.1(5)   \\
$^3S_1$ &     5.435(2)  &    1.852(2)  &    -0.122(3) &     1.429(7) &    -7.60(3)  \\
        &     5.42(1)   &    1.84(1)   &    -0.14(1)  &     1.46(3)  &    -7.7(2)   \\
$\EP_1$ &     1.630(6)  &    0.400(3)  &    -0.266(5) &     1.47(1)  &    -7.28(2)  \\
        &     1.61(2)   &    0.39(2)   &    -0.29(3)  &     1.47(2)  &    -7.35(9)  \\
$^3D_1$ &     6.46(1)   &   -3.540(8)  &    -3.70(2)  &     1.14(2)  &    -2.77(2)  \\
        &     6.43(4)   &   -3.57(2)   &    -3.77(4)  &     1.11(5)  &    -2.7(1)   \\
$^1D_2$ &    -1.376     &   15.04(2)   &    16.68(6)  &   -13.5(1)   &    35.4(1)   \\
        &    -1.379(6)  &   15.00(9)   &    16.7(2)   &   -12.9(4)   &    36.2(14)  \\
$^3D_2$ &    -7.400(4)  &    2.858(3)  &     2.382(9) &    -1.04(2)  &     1.74(2)  \\
        &    -7.39(1)   &    2.87(1)   &     2.41(3)  &    -0.96(5)  &     1.75(8)  \\
$^3P_2$ &    -0.290(2)  &   -8.19(1)   &    -6.57(5)  &    -5.5(2)   &   -12.2(3)   \\
        &    -0.288(5)  &   -8.3(2)    &    -6.8(7)   &    -6.1(19)  &   -12.7(26)  \\
$\EP_2$ &     1.609(1)  &  -15.68(2)   &   -24.91(8)  &   -21.9(3)   &   -64.1(7)   \\
        &     1.604(6)  &  -15.8(2)    &   -25.2(7)   &   -23.0(29)  &   -66.2(69)  \\
$^3F_2$ &    -0.971     &   -5.74(2)   &   -23.26(8)  &   -79.5(4)   &  -113.0(16)  \\
 &    -0.971(5)  &   -5.7(1)    &   -23.3(6)   &   -80.1(33)  &  -117.2(121) \\
 \end{tabular*}
}
\end{table}

\subsection{Low energy constants}

Alternatively, one may use effective interactions derived from a low momentum
interaction where the coefficients can be identified with the phenomenological
counter-terms of chiral effective field theory. To obtain such counter-terms
we express the momentum space NN potential in the partial wave basis
\begin{equation}
v^{JS}_{l',l} (p',p) =(4\pi)^2 \int_0^\infty \, dr\, r^2 \, j_{l'}(p'r)
j_{l}(pr) V_{l' l}^{JS}(r) \,
\end{equation}
and use the Taylor expansion of the spherical Bessel function 
to get an expansion for the potential in each partial wave. Keeping terms up
to fourth order ${\mathcal O} (p^4, p'^4, p^3p', pp'^3, p^2p'^2)$ corresponds
to keeping only $S$-, $P$- and $D$-waves along with $S$-$D$ and $P$-$F$ mixing
parameters.  Using the normalization and spectroscopic notation of
Ref.~\cite{Epelbaum:2004fk} one gets
\begin{align}
v_{00}^{JS}(p',p) 
=& 
 \widetilde{C}_{00}^{JS} 
+ C_{00}^{JS}(p^2+p'^2) 
+ D^1_{00}{}^{JS} (p^4+p'^4) 
+ D^2_{00}{}^{JS} p^2 p'^ 2 
+ \cdots 
\nonumber \\ 
v_{11}^{JS}(p',p) 
=& 
p p' C_{11}^{JS} 
+ p p' (p^2+p'^2) D_{11}^{JS} 
+ \cdots 
\nonumber \\
v_{22}^{JS}(p',p) 
=& 
p^2 p'{}^2 D_{22}^{JS} 
+ \cdots 
\nonumber   \\ 
v_{20}^{JS}(p',p) 
=& p'{}^2 C_{20}^{JS} 
+ p'{}^2 p^ 2  D^1_{20}{}^{JS} + p'{}^4 D^2_{20}{}^{JS} 
+ \dots \nonumber   \\
v_{31}^{JS}(p',p) 
=& 
p'{}^3 p D_{31}^{JS} 
+ \cdots 
\label{eq:Cts}
\end{align}
and each counter-term can be expressed as a radial momentum of the NN
potential in a specific partial wave. Different methods have been proposed to
quantify some of the uncertainties in these
quantities\cite{Epelbaum:2014efa,Furnstahl:2015rha}. Using the statistical
uncertainties method and the corresponding systematic error
estimates~\cite{Perez:2016vzj}, the results are summarized in
Table~\ref{tab:Cterms} for the 6 Granada potentials. 

\begin{table}[pt]
\caption{Potential integrals in different partial waves. Errors quoted for
  each potential are statistical; errors in the last column are
  systematic and correspond to the sample standard deviation of the
  six previous columns. See main text for details on the calculation
  of systematic errors. Units are: $\widetilde{C}$'s are in $10^4$
  ${\rm GeV}^{-2}$, $C$'s are in $10^4$ ${\rm GeV}^{-4}$ and $D$'s are
  in $10^4$ ${\rm GeV}^{-6}$. \label{tab:Cterms}}
    {\small \begin{tabular}{@{}cccccccc@{}} \toprule &DqS-OPE &DS-$\chi$TPE&
        DS-Born & Gauss-OPE &Gauss-$\chi$TPE&Gauss-Born &
        Compilation\\ \hline $\widetilde{C}_{^1S_0}$ & -0.141(1)&
        -0.135(2) & -0.128(2)& -0.121(5)& -0.113(9) & -0.133(3)&
        -0.13(1) \\ $C_{^1S_0}$ & 4.17(2) & 4.12(2) & 4.04(1) &
        4.20(2) & 4.16(2) & 4.18(1) & 4.15(6) \\ $D_{^1S_0}^1$
        &-448.8(11) & 443.7(5) &-441.5(3) &-447.0(10) & -446.7(2)
        &-446.3(2) &-445.7(26) \\ $D_{^1S_0}^2$ &-134.6(3) &-133.1(1)
        &-132.46(4) &-134.1(3) & -134.02(7) &-133.90(7) &-133.7(8)
        \\ $\widetilde{C}_{^3S_1}$ & -0.064(2)& -0.038(1) & -0.039(1)&
        -0.070(2)& -0.019(6) & -0.038(4)& -0.045(19)\\ $C_{^3S_1}$ &
        3.79(1) & 3.55(1) & 3.52(1) & 4.09(2) & 3.785(9) & 3.724(9)&
        3.7(2) \\ $D_{^3S_1}^1$ &-510.7(3) &-504.7(4) &-504.1(2)
        &-516.7(6) & -509.7(1) &-508.2(1) &-509.0(46) \\ $D_{^3S_1}^2$
        &-153.2(1) &-151.4(1) &-151.22(6) &-155.0(2) & -152.90(3)
        &-152.47(3) &-152.7(14) \\ $C_{^1P_1}$ & 6.44(2) & 6.54(1) &
        6.464(6)& 6.37(2) & 6.529(7) & 6.488(7)& 6.47(6)
        \\ $D_{^1P_1}$ &-594.9(2) &-592.1(2) &-590.21(6) &-594.5(2) &
        -597.83(7) &-596.25(7) &-594.3(28) \\ $C_{^3P_1}$ & 3.738(2)&
        3.659(3) & 3.633(3)& 3.762(6)& 3.677(3) & 3.599(1)& 3.68(6)
        \\ $D_{^3P_1}$ &-253.29(5) &-249.8(2) &-249.62(7) &-254.23(9)
        & -251.0(2) &-251.06(2) &-251.5(19) \\ $C_{^3P_0}$ &
        -4.911(8)& -4.882(5) & -4.897(3)& -4.944(6)& -4.802(8) &
        -4.883(2)& -4.89(5) \\ $D_{^3P_0}$ & 347.0(2) & 343.6(2) &
        344.62(6) & 345.8(1) & 345.02(3) & 346.25(2) & 345.4(12)
        \\ $C_{^3P_2}$ & -0.445(2)& -0.434(3) & -0.426(2)& -0.426(2)&
        -0.448(1) & -0.427(1)& -0.43(1) \\ $D_{^3P_2}$ & -10.62(7) &
        -9.7(2) & -9.45(6) & -11.55(4) & -9.939(8) & -9.631(7)&
        -10.1(8) \\ $D_{^1D_2}$ & -70.92(3) & -70.66(6) & -70.52(3) &
        -70.58(3) & -71.109(7) & -71.074(5)& -70.8(3) \\ $D_{^3D_2}$
        &-367.8(2) &-364.39(7) &-364.54(4) &-367.19(8) & -367.10(2)
        &-366.99(1) &-366.3(15) \\ $D_{^3D_1}$ & 205.8(2) & 204.25(7)
        & 204.26(4) & 204.4(1) & 205.17(3) & 205.21(3) & 204.9(6)
        \\ $D_{^3D_3}$ & 0.55(1) & 0.87(6) & 0.90(4) & -0.32(9) &
        0.26(3) & 0.51(3) & 0.46(45) \\ $C_{\epsilon_1}$ & -8.36(2) &
        -8.500(4) & -8.492(4)& -8.35(1) & -8.404(4) & -8.399(5)&
        -8.42(7) \\ $D_{\epsilon_1}^1$ &1012.6(6) &1005.5(1)
        &1006.23(6) &1010.5(3) & 1011.83(5) &1012.71(6) &1009.9(32)
        \\ $D_{\epsilon_1}^2$ & 434.0(3) & 430.94(4) & 431.24(3) &
        433.1(1) & 433.64(2) & 434.02(2) & 432.8(14)
        \\ $D_{\epsilon_2}$ & 84.18(4) & 83.29(1) & 83.398(7)&
        84.25(3) & 83.660(5) & 83.818(8)& 83.8(4) \\ \botrule
\end{tabular}}
\end{table}

\subsection{Scale dependence and correlations}

While one normally uses a {\it fixed} value for the maximum energy in the fits
(which in most NN studies has been 350 MeV), one may analyze the consequences
of varying this fitting energy~\cite{NavarroPerez:2012qr}. Denoting $\Lambda$
as the (running) maximal momentum it is clear that the fitting potential will
change as $\Lambda$ is varied. Actually, these parameters may be
mapped~\cite{Perez:2014kpa} into the so-called counter-terms which characterize
the effective theories at small momenta~\cite{RuizArriola:2016vap}. We
determined the two-body Skyrme force parameters arising from the NN
interaction as a function of the maximal momentum in the fit. We found general
agreement with the so-called $V_{\rm lowk}$ interactions based on high quality
potentials after high energy components have been integrated
out~\cite{Bogner:2003wn, Bogner:2009bt}. 

In line with our remarks in Section~\ref{sec:corr} let us note that,
one major outcome of Ref.~\cite{Perez:2014kpa} has been the fact that
the counter-terms corresponding to volume integrals including OPE
above 3 fm are weakly correlated, whereas those including OPE+TPE
above 1.8 fm have larger but still moderate correlations. Thus,
counter-terms in the partial waves basis would be efficient fitting
parameters, unlike in the cartesian basis. As we have already
discussed, using uncorrelated fitting parameters has the practical
consequence of reducing the computational determination of the least
squares minimization.

\section{Chiral vs non-chiral potentials}

In common with the analysis presented in the previous sections, much of the
early work on phase-shift analysis was undertaken long before the advent of
QCD, so the NN potentials were at most considered to be derivable from Quantum
Field Theory in purely hadronic terms.  This implies in particular the
One-Pion-Exchange potential, which has survived over the years, and the
Two-Pion-Exchange which has been changing depending on the computational
scheme since the first attempts in the early 50's (see e.g.
Ref.~\cite{Machleidt:1989tm} for a historical review, in particular about the
meson exchange picture).

After the appearance of QCD as a fundamental theory of strong interactions
there emerged dedicated studies on the underlying quark dynamics in terms of
quark cluster models, particularly concerning the origin of the nuclear core
(see e.g. Refs.~\cite{Oka:1984yw, AlvarezEstrada:1986wq, Valcarce:2005em} and
references therein). Despite the numerous attempts it is fair to say that
these investigations did provide some microscopic and quantitative
understanding of the short range components of the interaction but did not
offer an alternative to the conventional partial wave analysis.  Current QCD
potentials determined on the lattice~\cite{Aoki:2009ji, Aoki:2011ep,
Aoki:2013tba, Walker-Loud:2014iea}, are still less precise than
phenomenological ones.

In the early 90's Weinberg~\cite{Weinberg:1990rz} (see e.g.
\cite{Bedaque:2002mn, Epelbaum:2008ga, Machleidt:2011zz} for
comprehensive reviews and references therein) proposed an Effective
Field Theory (EFT) approach to NN scattering based on chiral symmetry
directly inspired by QCD features, where the spontaneous breakdown of
chiral symmetry underlies the would-be Goldstone boson nature of the
pion. As compared to the phenomenological approaches, the attractive
pattern of such an EFT was also the natural hierarchy of n-body forces
and the possibility of making an {\it a priori} estimate of the
systematic uncertainties in terms of a power counting to different
orders. This happened at about the time when the phenomenological
approach harvested its great success when the Nijmegen group obtained
for the first time a statistically acceptable $\chi^2/\nu \sim 1$ by
fitting and selecting np+pp scattering data. Comprehensive fits to
data with chiral interactions have been made using the N2LO chiral
potentials~\cite{Kaiser:1997mw} to the Nijmegen
database~\cite{Stoks:1993tb} for pp~\cite{Rentmeester:1999vw} and for
pp+np~\cite{Rentmeester:2003mf} and the N3LO chiral
potential~\cite{Entem:2002sf} to the enlarged
database~\cite{Machleidt:2000ge} for np~\cite{Entem:2003ft}. The
newest generation of chiral potentials have already provided fits to
the Granada-2013 database~\cite{Perez:2013oba, Perez:2013cza,
  Perez:2014bua, Piarulli:2014bda, Carlsson:2015vda, Piarulli:2016vel,
  Reinert:2017usi, Entem:2017gor}.

\subsection{Statistical issues}

Very recently chiral potentials to sixth order in the chiral expansion have been
been claimed by the Bochum group to outperform the non-chiral potentials on
the basis of the Granada-2013 database~\cite{Reinert:2017usi}. This was a
major achievement of the chiral approach (see also \cite{Entem:2017gor} for a
momentum space approach of the Idaho-Salamanca group). Another great advantage
of the chiral approach is that the number of fitting parameters is
substantially smaller than in the phenomenological approach. In no case,
however, have the authors taken seriously the available statistical tests to
verify {\it a posteriori} the normality of residuals.

Within the uncertainty quantification context, a critical analysis
with an eye on the future developments has been put forward in
Ref.~\cite{RuizArriola:2016sbf, NavarroPerez:2019sfj}. It has been
suggested that a further order in the expansion, namely N5LO, might
quite likely achieve the desired statistical consistency.  At the
present state, however, there are still some pending, hopefully
manageable, issues which need to be resolved before the validation of
the chiral approach to NN scattering can be declared without
reservations.

\subsection{The chiral tensorial structure}

For instance, the tensorial structure of the force requires
phenomenologically that all allowed NN components should contribute to
some extent to the total NN potential. Chiral perturbation theory
proposes a hierarchy among the different components so that the chiral
$W_Q$ component vanishes to N4LO, unlike all the phenomenological
analyses so far~\cite{NavarroPerez:2019sfj}.  In addition, the number
of independent parameters in a scheme where $W_Q$ would be
non-vanishing becomes  comparable to the phenomenological
potentials.

\subsection{Peripheral waves}

One of the reasons why the coupling constants discussed in Section
\ref{sec:yukawa} can be pinned down so accurately~\cite{Perez:2016aol,
Arriola:2016hfi} is given by the fact that long distant physics is rather well
determined. From that point of view one expects that peripheral waves are
rather sensitive to the shape of the potential and hence become independent of
the short range components. This also provides a method to validate other
analyses and in particular chiral potentials. A very vivid way of presenting
the discrepancy is by comparing the phase-shifts in terms of the impact
parameter variable~\cite{RuizSimo:2017anp} (see Eq. \ref{impact})  for every
partial wave
\begin{equation}
\xi^{\rm N4LO} (b) = \frac{\delta_l^{\rm N4LO} - {\rm Mean}(\delta_l)}{{\rm Std}(\delta_l)} \Big|_{l+1/2=b p} \, , 
\end{equation}
which provides a measure of the discrepancy with respect to a set of
phase-shifts (see Fig.~\ref{fig:errors-ps} for a plot of different
sets). The conclusion of ~\cite{RuizSimo:2017anp} is quite
unequivocal: In the range 2 fm $ \le b \le $ 5 fm the $\delta^{\rm
  N4LO}$ differ by more than $3\sigma$ when compared to the primary
Granada 2013 analysis for $F$, $G$ and $H$ waves, and become $1\sigma$
compatible with the spread of the 13 high quality potentials.

\subsection{Perturbation theory for higher partial waves}

The long distance character of chiral potentials suggests that one may
determine the high peripheral partial waves in perturbation theory, as done
explicitly in Ref.~\cite{Entem:2014msa}.  Actually, the low energy parameters
discussed above in Section~\ref{sec:lep} probe the longest distance features
of a given partial wave. Going to N2LO one sees that, while there is some
rough agreement between the perturbative and the full low energy parameters,
the detailed comparison including both statistical and systematic errors do
not agree.  Using the perturbative version of the variable phase approach, a
perturbative evaluation \cite{NavarroPerez:2019sfj} in the context of chiral
TPE (N2LO in the chiral expansion) was also undertaken and shown {\it not} to
converge to the exact result within uncertainties, even at the largest angular
momenta and hence for the most peripheral waves.

\subsection{Coarse graining chiral potentials}

Chiral potentials can be combined with coarse graining in a
statistically consistent way~\cite{Perez:2013cza, Perez:2013oba,
  Perez:2013cza, Perez:2014bua}. This allows for a reduction of
parameters to about 30 since the separation distance can be made as
small as $r_c=1.8$ fm without spoiling the statistical analysis. This
approach assumes the chiral power counting for the potential {\it
  above} $r_c$ but not in the coarse grained region so that the all
the potential components (including the chirally missing $W_Q$) are
non-vanishing, and taking $f^2=0.0075$ has provided natural values for
the chiral constants $(c_1, c_3, c_4)=(-0.41 \pm 1.08,-4.66 \pm 0.60,
4.31 \pm 0.17) {\rm GeV}^{-1} $ for $T_{\rm LAB} \le 350 {\rm MeV}
$~\cite{Perez:2013cza, Perez:2013oba}.

In contrast, the canonical (Weinberg) power counting scheme applies to
the {\it full} potential and only to at least N5LO provides all
non-vanishing tensorial components ($W_Q=0$ at N4LO), in which case
the number of parameters becomes comparable with the phenomenological
approach. As emphasized in Ref.~\cite{NavarroPerez:2019sfj}, the end
of the chiral road-map in NN scattering based on the power counting
will definitely occur when such a scheme becomes reliable enough to
select and fit scattering data, without explicit reference to the
phenomenological approach.

\section{Binding in Light nuclei: Error propagation}

Much of the previous analysis may be used to analyze the impact of NN
scattering uncertainties to binding energies.  A precursor of this type of
calculations was carried out in Ref.~\cite{Adam:1993pya} where estimates on
binding uncertainties were carried out using a statistical regularization of
phases and a direct solution of the inverse scattering problem.

\subsection{On-shell vs off-shell}

NN Scattering data describe only the behavior of nucleons on-shell, i.e. with
$E_p=\sqrt{p^2+M^2}$ in the relativistic case. However, nuclear structure
calculations usually need also the corresponding off-shell components so that
when going from the NN scattering data to the binding energy calculation some
extra information would be needed~\cite{srivastava1975off}. This ambiguity can
be used in fact to our benefit, since ideally one would determine the
off-shellness from the determination of the finite nuclei properties. The
successful attempts by  Vary {\it et al} are a good demonstration of
that~\cite{Shirokov:2005bk, Johnson:2010cd}

\subsection{Computational vs Physical precision}

Let us review the sources of numerical precision in the solution of the
quantum-mechanical problem. In the simplest NN case, where we
usually solve numerically the two-body Schr\"odinger equation, the precision
is fixed by the precision in the wave function. In the positive energy
situation corresponding to a scattering state we are rather interested in the
determination of the scattering phase-shifts

Within the few-body community there has been a trend to determine the
quantum mechanical solution with an increasing pre-defined precision,
say, a $1\%$.  This is a pure conventional precision which has been a
goal {\it per se} and, of course, good precision is not disturbing
provided the computational cost does not scale up to an unbearable
limit where the calculation becomes unfeasible.  However, this does
not correspond to the {\it physical} precision where all necessary
effects are taken into account and which determines in fact the
predictive power of the theory.

\subsection{Monte Carlo method}

The normality property of the residuals has been exploited to extract the
effective interaction parameters and corresponding
counter-terms~\cite{Perez:2014kpa} and to replicate via Monte Carlo bootstrap
simulation as a mean to gather more robust information on the uncertainty
characteristics of fitting parameters~\cite{Perez:2014jsa}.  We stress that
the verification of normality, Eq.~(\ref{eq:generator}), is essential for a
meaningful propagation of the statistical error, since the uncertainty
inherited from the fitted scattering data $\Delta O_i^{\rm exp}$ corresponds
to a genuine statistical fluctuation.  This allows to determine the $1\sigma$
error of the parameters ${\bf p}= {\bf p}_0 \pm \Delta {\bf p}^{\rm stat}$ and
hence the error in the potential
\begin{equation}
V_{NN} = V_{NN} ( {\bf p}_0 ) \pm \Delta V_{NN}^{\rm stat} 
 \label{eq:potential-we}
\end{equation}
which generates in turn the error in the NN phase-shifs $\delta = \delta ({\bf
p_0}) \pm \Delta \delta^{\rm stat}$ and mixing angles. Once the NN-potential
is determined the few body problem can be solved for the  binding energy, 
\begin{equation}
  \left[\sum_i T_i + \sum_{i<j} V_{NN}(ij) \right] \Psi = E_A \Psi
  \label{eq:triton}
\end{equation}
where 
\begin{equation}
E_A = E_A ({\bf p}_0) \pm \Delta E_A^{\rm stat} \, . 
\end{equation}

Direct methods to determine $\Delta {\bf p}^{\rm stat}$, $\Delta V_{NN}^{\rm
stat}$ and $\Delta E_A^{\rm stat} $ proceed either by the standard error
matrix or Monte Carlo methods (see e.g.~\cite{Nieves:1999zb}). In
Ref.~\cite{Perez:2014jsa} we have shown that the latter method is more
convenient for large number of fitting parameters (typically $N_P=40-60$), and
consists of generating a sufficiently large sample drawn from a multivariate
normal probability distribution
\begin{equation}
\label{eq:multinormal}
 P(p_1,p_2,\ldots,p_P) = \frac{1}{\sqrt{(2 \pi)^{N_P} \det {\mathcal E}}}
 e^{-\frac{1}{2}({\bf p}- {\bf p}_0)^T {\mathcal E}^{-1} ({\bf p}- {\bf
 p}_0)},
\end{equation}
where ${\mathcal E}_{ij}= (\partial^2 \chi / \partial p_i \partial p_j)^{-1}$
is the error matrix. We generate $M$ samples ${\bf p}_\alpha \in P $ with
$\alpha=1, \dots , M$, and compute $V_{NN}( {\bf p_\alpha})$ from which the
corresponding scattering phase shifts $\delta ({\bf p}_\alpha)$ and binding
energies $E_A ( {\bf p_\alpha})$ can be determined. Of course, one drawback of
the MonteCarlo propagation method is that the object function, in this case
the energy, needs to be evaluated a sufficiently large number of times which
may be unduly time consuming. An analysis of statistical errors at the phase
shift level shows that $M=25$ may be sufficient to reproduce consistently the
covariance matrix uncertainties from the MonteCarlo method.

\subsection{The deuteron}

The deuteron is the simplest bound nuclear np system for which the theory has
long been developed~\cite{Blatt:1952ije}.  Its quantum numbers $J^{P}=1^+$
correspond to the coupled $^3S_1-^3D_1$ channel with reduced wave functions
$u(r)$ and $w(r)$ respectively, so that we solve the bound state problem with
$E_d=-B_d= -\gamma^2/ 2\mu_{np}$, i.e. with $p=i \gamma$.  At long distances
\begin{equation}
  u(r) \to A_S e^{-\gamma r} \, ,  \qquad   
w(r) \to \eta A_S e^{-\gamma r} \left[1 + \frac{3}{\gamma r}+ \frac{3}{(\gamma r)^2} \right]
\end{equation}
For normalized states we list in Table~\ref{tab:DeuteronP} the asymptotic D/S
ratio $\eta$, asymptotic S-wave amplitude $A_S$, mean squared matter radius
$r_m$, quadrupole moment $Q_D$, D-wave probability $P_D$ and inverse matter
radius $\langle r^{-1} \rangle$ for some high quality potentials compared with
two Granada potentials, DS-OPE~\cite{Perez:2013mwa},
DS-TPE~\cite{Perez:2013oba}. The PWA analysis indeed uses its binding energy
as a fitting parameter, so that the quoted uncertainties are purely
statistical. Unlike $r_m$, $Q_D$ or $P_D$ which require (small) meson exchange
currents corrections before being compared to experimental data, $A_S$ and
$\eta$ are purely hadronic. As we see, both the DS-OPE~\cite{Perez:2013oba}
DS-TPE~\cite{Perez:2013oba} provide {\it smaller} uncertainties than the
experimental/recommended values for $A_S$ and $\eta$. To our knowledge, this
is an unprecedented situation in Nuclear Physics. Similar trends are also
observed for the corresponding deuteron charge, magnetic and quadrupole form
factors (see e.g. \cite{Gilman2002} for a review) where DS-OPE
\cite{Perez:2013mwa} and DS-TPE~\cite{Perez:2013oba, Perez:2013za} generate
tiny uncertainties and offer an opportunity to discriminate meson exchange
currents contributions.

\begin{table}[ht]
\small \centering
	\caption{Deuteron static properties compared with
          empirical/recommended
          values~\cite{VanDerLeun,Borbely,Rodning:1990zz,Klarsfeld1986373,Bishop:1979zz,deSwart:1995ui}
          and high-quality potentials calculations,
          DS-OPE~\cite{Perez:2013mwa}, DS-TPE~\cite{Perez:2013oba},
          Nijm I~\cite{Stoks:1994wp}, Nijm II~\cite{Stoks:1994wp},
          Reid93~\cite{Stoks:1994wp}, AV18~\cite{Wiringa:1994wb},
          CD-Bonn~\cite{Machleidt:2000ge}.  We list binding energy
          $E_d$, asymptotic D/S ratio $\eta$, asymptotic S-wave
          amplitude $A_S$, mean squared matter radius $r_m$,
          quadrupole moment $Q_D$, D-wave probability $P_D$ and
          inverse matter radius $\langle r^{-1} \rangle$.}
	\label{tab:DeuteronP}
	\begin{tabular*}{\columnwidth}{@{\extracolsep{\fill}}l l l l l l l l l}
      \hline
            & Emp./Rec. & DS-OPE & DS-TPE & Nijm I  & Nijm II & Reid93  & AV18 & CD-Bonn  \\
		$E_d$(MeV)               & 2.224575(9)    & Input       & Input       & Input    & Input    & Input    & Input  & Input  \\
		$\eta$                   & 0.0256(5)      & 0.02493(8)  & 0.02473(4) & 0.02534  & 0.02521  & 0.02514  & 0.0250 & 0.0256 \\
		$A_S ({\rm fm}^{1/2})$     & 0.8845(8)     & 0.8829(4)  & 0.8854(2)  & 0.8841   & 0.8845   & 0.8853   & 0.8850 & 0.8846 \\
		$r_m ({\rm fm})$          & 1.971(6)       & 1.9645(9) & 1.9689(4)   & 1.9666   & 1.9675   & 1.9686   & 1.967  &  1.966 \\
		$Q_D ({\rm fm}^{2}) $     & 0.2859(3)      & 0.2679(9)  & 0.2658(5)  & 0.2719   & 0.2707   & 0.2703   & 0.270  & 0.270  \\
		$P_D$                      & 5.67(4)        & 5.62(5)  & 5.30(3)     & 5.664    & 5.635    & 5.699    & 5.76   & 4.85   \\
		$\langle r^{-1} \rangle ({\rm fm}^{-1})$ &  & 0.4540(5) & 0.4542(2)  &          & 0.4502   & 0.4515   &        &        \\
      \hline \hline
	\end{tabular*}
\end{table}

\subsection{Binding energies for A=3,4 systems}

The primary Granada DS-OPE potential which was used to fit and select np+pp
scattering data uses Dirac delta-shells which are too singular in
configuration space or have too long momentum tails, for instance in the
deuteron~\cite{Perez:2015bqa}, to be handled in few body calculations.
Actually, this was the reason to design smooth SOG (Sum of Gaussian)
potentials~\cite{Perez:2014yla, Perez:2014waa} referenced in
Section~\ref{sec:6Gr}.

In Ref.~\cite{Perez:2014laa} the triton binding energy was evaluated
for the SOG-OPE Granada potential using the hyper-spherical harmonics
method with $M \sim 200$ MonteCarlo replicas, and statistical
distributions where also obtained yielding $\Delta E_t = 12$KeV. One
motivation for such a calculation was to determine if the
computational accuracy was unnecessarily better than the statistical
accuracy inherited from the NN scattering data. Our points are
illustrated in Table~\ref{tab:Etriton} from Ref.~\cite{Perez:2014laa}
where the numerical convergence regarding the number of partial waves
is displayed. The error estimate clearly marks where the accuracy of
the numerical calculation is larger than the physical accuracy.

\begin{table}[th]
	\centering
        \caption{Triton binding energy convergence for the
          hyper-spherical harmonics method~\cite{Perez:2014laa} in the
          number of channels, $N_c$, classified according to the
          orbital angular momentum of the pair $L_{\rm Pair}$ and the
          spectator $l_{\rm spectator}$ in the triton as the number of
          total accumulated channels, $N_{\rm Total}$, is increased.
          The potential used was Monte Carlo generated. A horizontal
          line is drawn when the change in $E_t$ is {\it smaller} than
          the statistical uncertainty $\Delta B_t = 15(1) \, {\rm
            keV}$.}
      \label{tab:Etriton}
	\begin{tabular*}{\columnwidth}{@{\extracolsep{\fill}}c c c c }
            \hline
            \hline\noalign{\smallskip}
             $N_c$ & $L_{\rm Pair} \, l_{\rm Spectator}$ &  $N_{\rm Total}$ & Energy (MeV)\\
            \hline\noalign{\smallskip}
             3 & Ss  & 3 &  Unbound \\ 
             +2  & Sd+Ds & 5 & -7.0117  \\ 
             +10  & Pp    & 15 & -6.4377 \\ 
             +8 & Dd    & 23 & -7.4109  \\ 
             +4  & Pf+Fp & 27 & -7.4956  \\ 
             +10  &  Ff   & 37 & -7.5654  \\ 
             +2  & Dg+Gd  & 39 &   -7.6178 \\ 
             +8  & Gg  &  47 &  -7.6502 \\ \hline 
             +4  & Fh+Hf  & 51 &   -7.6508 \\ 
             +10  & Hh  &  61 &  -7.6510 \\      
            \noalign{\smallskip}\hline
            \hline
	\end{tabular*}
\end{table}


The statistical uncertainty of experimental NN scattering data have also been
propagated into the binding energy of $^3$H and $^4$He using the no-core full
configuration method in a sufficiently large harmonic oscillator basis.  The
error analysis~\cite{Perez:2015bqa} yields $\Delta B_t = 15$ KeV and $\Delta
B_\alpha = 55$ KeV.

Similar patterns occur when solving the Faddeev equations for $^3$H and the
Yakubovsky equations for $^4$He respectively~\cite{Perez:2016oia}. We check
that in practice about $M=30$ samples prove enough for a reliable error
estimate within the MonteCarlo method, giving $\Delta B_t = 12$KeV and $\Delta
B_\alpha = 50$KeV whereas, again, the computational accuracy is better,
$\Delta B_t^{\rm num}=1$ KeV and $\Delta B_\alpha^{\rm num}=20$ KeV .

Results for the 3N and 4N binding energies for various NN potentials
using the Faddeev equations for $^3$H and the Yakubovsky equations for
$^4$He are listed in Table~\ref{tab:3H-4He} where we see a systematic
underbinding with respect to the experimental values. A popular
interpretation of this disagreement suggests that the influence of
three- and four-body forces has been neglected. However, the
contribution of three body forces depends on the definition of two
body forces as we will discuss next.

\subsection{The Tjon line}

Much of the error analysis which can and has been carried out in Nuclear
Physics is probably best exemplified by the so called Tjon
line~\cite{Tjon:1975sme}, a linear but empirical correlation between the
triton and $\alpha$-particle binding energies of the form
\begin{equation}
 B_\alpha = a B_t + c 
\end{equation}

where $a,c$ depend on a family of NN potentials which have the {\it same} NN
scattering phase shifts and deuteron properties. Thus, the slope may be
schematically be written as $ a = (\partial B_\alpha/\partial B_t)|_{B_d} $.
This empirical feature~\cite{Perne:1979zz, Tjon:1981lhu} comparing between
phase-equivalent potentials has been corroborated by many calculations ever
since~\cite{Nogga:2000uu, Nogga:2004ab, Klein:2018lqz}.  It is remarkable that
such a simple property has no obvious explanation. One clue would be the fact
that the deuteron binding energy, $B_d = 2.2$ MeV,  is small compared to the
triton and alpha bindings  \cite{Delfino:2007zu}. For small  $B_d$ the alpha
binding energy then would scale as $B_\alpha = a B_t + b B_d + {\mathcal O}(
B_d^2) $. The points along this line in the plane $( B_t, B_\alpha)$ 
correspond to potentials with the same phase-shifts, verifying  $\Delta
B_\alpha = a \Delta B_t $  The points along a perpendicular line,  $\Delta
B_\alpha = -1/a \Delta B_t $  should correspond to potentials with very
different phase-shifts.  In particular the difference may be generated by a
unitary transformation of the NN potential, $V_2 \to U V_2 U^\dagger$, so that
the bindings depend on $U$ but the coefficients $a$ and $b$ do not depend on
$U$~\cite{Nogga:2000uu}. On the other hand, a unitary transformation of the
{\it two-body} potential implies a change in multi-nucleon forces, $V_3$,
$V_4$, etc.  and, one may actually fit $E_t$ with a suitable $V_3$ and
$E_\alpha$ with a suitable $V_4$ yielding for $V_4=0$ in the so-called
on-shell limit the formula $B_\alpha = 4 B_t - 3 B_d$ which works
well~\cite{Arriola:2013gya, Arriola:2016fkr}.

\begin{table*}[tbp]
    \caption{3N and 4N binding energies for various NN potentials
      using the Faddeev equations for $^3$H and the Yakubovsky
      equations for $^4$He
      respectively~\cite{Nogga:2000uu,Perez:2016oia}. Errors in SOG-OPE
    are statistical.}
    \label{tab:3H-4He}
  \begin{center}
    \begin{tabular}{l|r|r|r|r|r|r|r} 
Potential &  Exp. & SOG-OPE & CD Bonn & AV18  & Nijm I & Nijm II & Nijm93  \\
\hline
$^3$H [MeV] & -8.4820(1)  & -7.660(12)  & -8.012 & -7.623  & -7.736   & -7.654 & -7.668 \\
$^4$He [MeV]& -28.2957(1) & -24.760(47)    & -26.26    &  -24.28    &   -24.98    &   -24.56 & -24.53 \\
\hline 
    \end{tabular}
  \end{center}
\end{table*}

Phase equivalent interactions produce a Tjon slope which is typically about
$\Delta B_\alpha/\Delta B_t \sim 5-6$ both in the
Faddeev-Yakubovsky~\cite{Nogga:2004ab} and in the no-core shell
model~\cite{Shirokov:2011dc}.  For the Faddeev-Yakubovsky solutions of
$^3$H-$^4$He the results from five high quality potentials, i.e. with
$\chi^2/\nu \sim 1$ at their time and the Granada SOG-OPE, in
Table~\ref{tab:3H-4He} give $B_\alpha = 4.73 B_t - 5.26 B_d$. For a sample of
SOG-OPE potentials the statistical bootstrap analysis with $M=30$ gives
$B_\alpha = 4.8(1) B_t - 5.4(3)B_d$, where the central values reflect the
actual scattering data and the uncertainties reflect the truly
phase-inequivalent fluctuations. The extrapolation predicts the experimental
binding of the alpha particle within uncertainties~\cite{Perez:2016oia}, since
\begin{equation}
\Delta B_\alpha^2|_{\rm stat} = (\Delta a)^2 B_t^2 +  (\Delta b) ^2 B_d^2 
\end{equation}
so that $\Delta B_\alpha|_{\rm stat} \sim 1 {\rm MeV}$.  Interestingly, this
suggests a marginal effect of four body forces, for which independent
estimates using approximate wave functions~\cite{Rozpedzik:2006yi} give
similar numbers, $B_\alpha|_{\rm 4N} \sim -100$KeV (see also
Ref.~\cite{Epelbaum:2007us} for a chiral scheme where this is argued to
overestimate the result.). Thus, we see that since $B_\alpha|_{\rm 4N} \sim
\Delta B_\alpha^{\rm stat}$ the four-body force might be unobservable. While
this is good news from the theoretical point of view, more detailed
calculations might be needed to confirm this feature. Finally, let us also
mention that along these lines, theoretical uncertainties of the elastic
nucleon-deuteron scattering observables have been
undertaken~\cite{Skibinski:2018dot}.

\section{Effective Nuclear Interactions}

\subsection{Moshinsky-Skyrme parameters}

Power expansions in momentum space of effective interactions were introduced
by Moshinsky \cite{moshinsky1958short} and Skyrme \cite{skyrme1958effective}
to provide significant simplifications to the nuclear many body problem in
comparison with the {\it ab initio} approach, in which it is customary to
employ phenomenological interactions fitted to NN scattering data to solve the
nuclear many body problem. As a consequence of such simplifications effective
interactions, also called Skyrme forces, have been extensively used in mean
field calculations \cite{Vautherin:1971aw, Negele:1972zp, Chabanat:1997qh,
Bender:2003jk}. Within this framework the effective force is deduced from the
elementary NN interaction and encodes the relevant physical properties in
terms of a small set of parameters. However, there is not a unique
determination of the Skyrme force and different fitting strategies result in
different effective potentials (see e.g. Refs.~\cite{Friedrich:1986zza}
and~\cite{Klupfel:2008af}). This diversity of effective interactions within
the various available schemes signals a source of statistical and systematic
uncertainties that remain to be quantified. Fortunately the parameters
determining a Skyrme force can be extracted from phenomenological interactions
\cite{Arriola:2010hj, NavarroPerez:2012qr} and uncertainties can be propagated
accordingly \cite{Perez:2014kpa}. At the two body level the Moshinsky-Skyrme
potential in momentum representation reads
\begin{align} 
 V_\Lambda ({\bf
    p}',{\bf p}) 
&= \int d^3 x e^{-i {\bf x}\cdot ({\bf p'}-{\bf p})}  \hat V({\bf x} ) 
 \nonumber \\ &=  t_0 (1 + x_0 P_\sigma ) + \frac{t_1}2(1 + x_1
  P_\sigma ) ({\bf p}'^2 + {\bf p}^2) \nonumber \\ 
&+  
 t_2 (1 + x_2
  P_\sigma ) {\bf p}' \cdot {\bf p} + 2 i W_0 {\bf S} \cdot({\bf p}'
  \times {\bf p}) \nonumber \\ &+ 
\frac{t_T}2 \left[ \sigma_1 \cdot {\bf p}
  \, \sigma_2 \cdot {\bf p}+ \sigma_1 \cdot {\bf p'} \, \sigma_2
  \cdot {\bf p'} - \frac13 \sigma_1 \, \cdot 
\sigma_2 ({\bf p'}^2+  {\bf p}^2)
\right] \nonumber \\  &+
\frac{t_U}2 \left[ \sigma_1 \cdot {\bf p}
  \, \sigma_2 \cdot {\bf p}'+ \sigma_1 \cdot {\bf p'} \, \sigma_2
  \cdot {\bf p} - \frac23 \sigma_1 \, \cdot 
\sigma_2 {\bf p'}\cdot  {\bf p}
\right]  
+ {\mathcal O} (p^4) 
\label{eq:skyrme2}
\end{align} 
where $P_\sigma = (1+ \sigma_1 \cdot \sigma_2)/2$ is the spin exchange
operator with $P_\sigma=-1$ for spin singlet $S=0$ and $P_\sigma=1$ for spin
triplet $S=1$ states. These parameters correspond to radial moments of volume
integrals of the potentials $ \int_0^\infty d^3 x r^n V_i(r) $ which are
increasingly insensitive to short distances.

\begin{table}[ht]
\caption{Moshinsky-Skyrme parameters for the renormalization scale
  $\Lambda=400$ MeV. Errors quoted for each potential are statistical;
  errors in the last column are systematic and correspond to the
  sample standard deviation of the six previous columns. See main text
  for details on the calculation of systematic errors. Units are:
  $t_0$ in ${\rm MeV} {\rm fm}^3$, $t_1,t_2,W_0,t_U,t_T$ in ${\rm MeV}
  {\rm fm}^5$, and $x_0,x_1,x_2$ are
  dimensionless. \label{tab:Skyrme}}
    {\small \begin{tabular}{@{}cccccccc@{}}
    \toprule & DS-OPE & DS-$\chi$TPE& DS-Born & Gauss-OPE
    &Gauss-$\chi$TPE& Gauss-Born & Compilation \\ \hline $t_0$ &
    -626.8(64) & -529.6(53) & -509.0(55) & -584.4(157) & -406.1(289) &
    -521.8(152) & -529.6(751) \\ $x_0$ & -0.38(2) & -0.56(1) &
    -0.54(1) & -0.26(2) & -0.71(8) & -0.55(4) & -0.50(16) \\ $t_1$ &
    948.1(30) & 913.6(22) & 900.1(17) & 987.4(29) & 945.5(18) &
    941.3(16) & 939.3(304) \\ $x_1$ & -0.048(3) & -0.074(3) &
    -0.068(3) & -0.013(3) & -0.047(3) & -0.058(2) & -0.051(22)\\ $t_2$
    & 2462.6(56) & 2490.0(39) & 2462.1(25) & 2441.3(56) & 2490.1(24) &
    2466.8(26) & 2468.8(187) \\ $x_2$ & -0.8686(6)& -0.8750(8)&
    -0.8753(6)& -0.8630(8)& -0.8729(6) & -0.8785(3)& -0.872(6)
    \\ $W_0$ & 107.7(4) & 100.8(3) & 96.2(3) & 105.0(5) & 109.3(7) &
    94.3(2) & 102.2(61) \\ $t_U$ & 1278.6(12) & 1260.3(5) & 1257.0(4)
    & 1285.6(12) & 1254.9(9) & 1249.3(3) & 1264.3(144) \\ $t_T$
    &-4220.9(87) &-4292.8(23) &-4289.0(21) &-4385.6(99) & -4271.8(51)
    &-4319.5(58) & -4296.6(545) \\ \botrule
\end{tabular}}
\end{table}

As mentioned above different nuclear data can be used to constrain the Skyrme
potential. The usual approach is to fit parameters of Eq.~(\ref{eq:skyrme2})
to doubly closed shell nuclei and nuclear matter saturation
properties~\cite{Vautherin:1971aw, Negele:1972zp, Chabanat:1997qh,
Bender:2003jk}. In Ref.~\cite{Arriola:2010hj} the parameters were determined
from just NN threshold properties such as scattering lengths, effective ranges
and volumes without explicitly taking into account the finite range of the NN
interaction; while in Ref.~\cite{NavarroPerez:2012qr} the parameters were
computed directly from a local interaction in coordinate space that reproduces
NN elastic scattering data. In Ref.~\cite{Perez:2014kpa} the latter approach
was used to propagate \emph{statistical} uncertainties into the Skyrme
parameters. The quantification of the \emph{systematic} uncertainties, which
arise from the different representations of the NN interaction was discussed
in Ref.~\cite{Perez:2016vzj}. The results, summarized in
Table~\ref{tab:Skyrme} clearly show, again, the dominance of systematic vs
statistical errors.

\subsection{Error estimates for heavy nuclei and nuclear matter}

Within the Skyrme effective interactions approach one can find a simple
estimate of systematic errors due to the two body interaction uncertainty
using (for a review see \cite{Bender:2003jk}) 
\begin{equation}
\frac{\Delta B}{A} =   
\frac{3}{8 A} \Delta t_0 \, \int d^3 x \, \rho (x)^2 \,   , 
\end{equation}
For nuclear matter at saturation, $\rho_0= 0.17 {\rm fm}^{-3}$, our $\Delta
t_0= 75 {\rm MeV}\, {\rm fm}^3$ implies
\begin{equation}
\frac{\Delta B}{A} = \frac{3}{8} \Delta t_0 \rho_0  = 2.4 {\rm MeV}  \, . 
\end{equation}
We may implement finite size effects in light-heavy nuclei by using a
Fermi-type shape for the matter density
\begin{equation}
\rho(r) =   C/(1+e^{(r-R)/a})
\end{equation}
with  $R=r_0 A^{\frac13}$,   $r_0=1.1 {\rm fm}$ and $a=0.7 {\rm fm}$, 
Normalizing to the total number of particles $A = \int d^3 x \rho(x)$ we get 
values in the range
\begin{equation}
\Delta B_A/A =  0.4-1.6 {\rm MeV} \, , 
\end{equation}
depending on the value of $A$ for $4 \le A \le 208$.

\section{Coarse grained potential results}

Besides the aspect of uncertainty quantification which is the focus of
the present work, we believe that the very idea of coarse graining
proves useful in nuclear physics. This requires that special methods
have to be developed for delta--shells interaction, which in our view
are the most flexible ones which allow for selecting and fitting the
largest NN database to date, but cannot be plugged directly in
conventional computing codes dealing with nuclear structure and
reactions, and hence smooth potentials (such as the SOG-Granada type
potentials) need to be defined {\it after} the data selection process.
This is similar to what happened with the energy dependence needed by
the Nijmegen group which also led to subsequent high quality
interactions.  We discuss here some simple examples where delta-shells
may be used {\it directly}. 

\subsection{Repulsive vs structural core}

Besides the well accepted OPE mechanism for long distances and the mid-range
attraction which is needed for nuclear binding, one of the traditional and
well accepted properties of the nuclear potential is the existence of a
nuclear strongly repulsive core at about $0.5$ fm. While this feature
guarantees the stability of nuclei and nuclear matter against collapse it also
complicates the solution of the many body problem, since the relative NN wave
function must vanish below the core, therefore introducing a very strong short
range correlation.  At a practical level the existence of the core implies a
vanishing of the wave function at about the core location, but something else
is needed to determine the wave function below the core radius. The question
is whether the repulsive core is indispensable from from the analysis of
collision experiments. However, in order to resolve the core in NN elastic
scattering one needs a wavelength which corresponds to energies where there is
a substantial in-elasticity and hence a complex optical potential is needed in
order to deal with the absorption due to inelastic processes such as $NN \to
NN \pi$. This point has been analyzed in Ref.~\cite{Fernandez-Soler:2017kfu}
and it has been found that there exist two solutions, one corresponding to the
usual repulsive core and the other one related to the so-called structural
core, reminiscent of the composite character of the nucleon. 

\subsection{Coarse graining short range correlations}

The Bethe-Goldstone equation \cite{Bethe:1956zz, Goldstone:1957zz} has been a
way to describe short range correlations between nucleons inside the nucleus.
In the nuclear medium the interaction produces no scattering due to the Pauli
principle. Instead the relative wave function of a pair is modified in
presence of the two-body interaction, generating high-momentum components
above the Fermi momentum, $p> p_F$. Using the delta-shell potential allows to
simplify the problem of computing these high momentum components arising in an
interacting nucleon pair in nuclear matter.  This coarse graining of the
Bethe-Goldstone equation has been explored in
~\cite{RuizSimo:2016vsh,RuizSimo:2017tcb} for back-to-back nucleons, with
total center of mass momentum equal to zero. The formalism still  has to be 
extended to other values of the center of mass.


\subsection{Error analysis of nuclear matrix elements}

The expected errors of harmonic oscillator nuclear matrix elements coming from
the uncertainty on the NN interaction have been estimated in
\cite{Amaro:2013zka} for the coarse grained (GR) interaction fitted to NN
scattering data, with several prescriptions for the long-part of the
interaction, including one pion exchange and chiral two-pion exchange
interactions.

\subsection{Shell model estimates}

In a previous calculation~\cite{NavarroPerez:2011fm}, we showed how our
approach is competitive not only as a way of determining the phase shifts but
also compared to more sophisticated approaches to Nuclear
Structure~\cite{Neff:2002nu}. We computed the ground state energy of several
closed-shell nuclei by using oscillator wave functions. In the case of $^4
{\rm He}$, $^{16}{\rm O}$ and $^{40}{\rm Ca}$ nuclei, our calculation
reproduces the experiment at the $20-30\%$-level provided the phase-shifts are
fitted up to energy $E \leq 100 {\rm MeV}$~\cite{NavarroPerez:2011fm}.  This
is a tolerable accuracy as we just intend to make a first estimate on the
systematic uncertainties and then compute the change in the binding energy.
For the $A=3,4$ nuclei we use the simple formulas,
\begin{align}
\Delta B(^3{\rm H}) 
=&
\langle \Delta V_2 \rangle_{^3{\rm H}} 
= 3 \langle 1s | \frac12 \left( \Delta V_{^1S_0} + \Delta V_{^3S_1} \right) | 1s \rangle  \, , 
\label{eq:db_3}\\ 
\Delta B(^4{\rm He}) 
=&
\langle \Delta V_2 \rangle_{^4{\rm He}} 
= 6 \langle 1s | \frac12 \left( \Delta V_{^1S_0} + \Delta V_{^3S_1} \right) | 1s \rangle \, , 
\label{eq:db_4} 
\end{align}
where $ |1s \rangle $ is the Harmonic oscillator relative wave function with
the corresponding oscillator parameter $b$ fixed to reproduce the physical
charge radius. The factors in front of the matrix elements are Talmi-Moshinsky
coefficients  corresponding in this particular case to the number of pairs
interacting through a relative s-wave. Errors in the potential $\Delta V$  are
computed by adding individual contributions $ (\Delta \lambda_n)^{JS}_{l,l'} $
in quadrature. By propagating the potential errors
to Eq.~(\ref{eq:db_3}) we find 
\begin{equation}
\frac{\Delta B(3)}{3} = 0.07-0.085 {\rm MeV}
\end{equation}
depending on the fitting cut-off LAB energy, 100-350 MeV respectively,
overestimating the Faddeev estimates given above. For the $\alpha-$particle
Eq.~(\ref{eq:db_4}) yields 
\begin{equation}
\frac{\Delta B(4)}{4} =
0.10-0.13 {\rm MeV}.
\end{equation}  
More generally, for heavier double-closed shell nuclei one has along the lines
of Ref.~\cite{NavarroPerez:2011fm}
\begin{equation}
\Delta B(A) = \sum_{nlSJ}g_{nlJS}
\langle nl | \Delta V^{JST}|nl\rangle
\end{equation}
where $g_{nlJS}$ depends on the Talmi-Moshinsky brackets. For $^{16}$O and
$^{40}$Ca, we find
\begin{equation}
\frac{\Delta B(^{16}{\rm O})}{16} = 0.26\, {\rm MeV} \qquad 
\frac{\Delta B(^{40}{\rm Ca})}{40} = 0.32 {\rm MeV}.
\end{equation}
These systematic estimates using shell model are of the same order to the ones
obtained above in the Skyrme interaction.

\section{Outlook}

Despite the many years elapsed since the first NN partial wave
analysis in 1957 and the huge theoretical and experimental efforts
carried out, the nuclear force is poorly known still where it is most
needed, namely in the mid-range regime which is relevant for {\it ab
  initio} calculation of nuclear binding energies. This is the
explanation behind the relatively large uncertainties found in large
scale calculations. During many years there has been a conformist
attitude regarding these uncertainties, and in most papers a purely
computational approach has prevailed, validating theoretical
frameworks just on their numerical performance. Only in recent years
the issue of uncertainties has been taken seriously, as it is actually
the key to establish the predictive power of the theory. Clearly, the
level of ambiguity we are dealing with in the evaluation of nuclear
uncertainties of all sorts, statistical , systematic and computational
requires a rigorous treatment. In this work we have reviewed this
topic from the perspective of the impact of the Granada NN database on
the determination of the NN force and its consequences on nuclear
binding.

The main theoretical obstacle has to do with the great difficulty in
providing a unique definition of the nuclear potential just from
data. Quantum field theory at the hadronic level implies the existence
of a long range interaction dominated by pion exchanges as the
lightest particles and reduces the ambiguity. Lattice calculations of
potentials may identify them with static energies assuming heavy
quark-composite sources but their accuracy is at present not
satisfactory. Chiral perturbation theory provides in addition several
schemes based on a power counting which, while not fully satisfactory,
may be and have been implemented in the NN sector and extended to
multi-nucleon forces. The consistency among chiral multi-nucleon
forces is theoretically very appealing and the use of potentials is
possibly the only practical path towards a satisfactory solution of
the nuclear many body problem. It should be stressed that the EFT
point of view is the most suitable one since in principle one gets rid
of the model dependence with {\it a priori} uncertainty estimates.
However, a more detailed analysis reveals that there are issues
regarding the necessary regularization of the theory, which
effectively model the mid-range regime of the NN interaction.
Moreover, the indispensability of the chiral scheme for NN scattering
data remains to be proven, not to speak about its suitability for
fitting and selecting a NN database itself.  At a phenomenological
level at the present stage the determination of the NN interaction
below 1.8 fm (up to a phase equivalent unitary transformation) remains
so far connected to a combination of an abundance of data in a variety
of kinematics and observables with the corresponding experimental
errors.

In our view, this unfortunate situation on the side of the hadronic
theory will likely not necessarily improve neither with more and
better experimental measurements nor with larger computational
facilities, but with a better understanding on the essence of hadronic
interactions and their range of applicability.

\section*{Acknowledgments}
We thank Ignacio Ruiz Sim\'o, James Vary, Pieter Maris, Eduardo
Garrido, Andreas Nogga, Rocco Schiavilla, Maria Piarulli, Pedro
Fernández Soler, Jacobo Ruiz de Elvira, Varese Salvador Timoteo and
Sergio Szpigel for collaboration on different issues discussed in this
paper. This work is supported by the Spanish MINECO and European FEDER
funds (grant FIS2017-85053-C2-1-P) and Junta de
Andalucía (grant FQM-225)



\end{document}